\documentclass[a4paper,notitlepage,11pt,onecolumn]{article}
 
\begin{document}

 \title{ Spin glass transition in a magnetic field: 
a renormalization group study}
 
 \author{
 I.R. Pimentel
 \\
 Department of Physics and CFMC, University of Lisbon,
 \\
 Av. Prof. Gama Pinto, 2, 1649 Lisboa, Portugal
 \\
 \bigskip
 \\
 T. Temesv\'{a}ri
 \\
 HAS Research Group for Theoretical Physics, E\"{o}tv\"{o}s University,
 \\
 H-1117 P\'{a}zm\'{a}ny P\'{e}ter s\'{e}t\'{a}ny 1/A, Budapest, Hungary
 \\
 \bigskip
 \\
 C. De Dominicis
 \\
 Service the Physique Th\'{e}orique, CEA Saclay,
 \\
 F-91191 Gif-sur-Yvette, France}

 \date{(Dated: February 12, 2002)}
 
 \maketitle
 
 \begin{abstract}
 We study the transition of short range Ising spin glasses in a magnetic
 field, within a general replica symmetric field theory, which contains three
 masses and eight cubic couplings, that is defined in terms of the fields
 representing the replicon, anomalous and longitudinal modes. We discuss the
 symmetry of the theory in the limit of replica number $n\rightarrow 0$, and
 consider the regular case where the longitudinal and anomalous masses remain
 degenerate.
  The spin glass transitions in zero and non-zero field are analyzed in
 a common framework. The mean field treatment shows the usual results, that
 is a transition in zero field, where all the modes become critical, and a
 transition in nonzero field, at the de Almeida-Thouless (AT) line, with only
 the replicon mode critical. Renormalization group methods are used to study
 the critical behavior, to order $\varepsilon =6-d$. In the general theory we
 find a stable fixed-point associated to the spin glass transition in zero
 field. This fixed-point becomes unstable in the presence of a small magnetic
 field, and we calculate crossover exponents, which we relate to zero-field
 critical exponents. In a finite magnetic field, we find no physical stable
 fixed-point to describe the AT transition, in agreement with previous
 results of other authors.
 \end{abstract}
 \bigskip
 \bigskip
 

 \renewcommand{\thesection}{\Roman{section}}

 \section{INTRODUCTION}

 Spin glasses are specially interesting disordered magnetic systems, with
 competing interactions which generate frustration.$^{1-4}$ Concepts and
 techniques developed in the study of these complex systems have had impact
 in a variety of other subjects, like combinatorial optimization, neural
 networks, prebiotic evolution and protein folding.$^{2,5}$ Here we
 concentrate on the simplest spin glass model, that is the Ising spin glass
 with a Gaussian distribution of interactions in a uniform magnetic field.
 Despite the enormous amount of work dedicated over the past twenty years, to
 the study of spin glasses, no consensus has yet been reached on the most
 fundamental properties of these systems, namely the nature and complexity of
 the glassy phase and the existence of a transition in a nonzero magnetic
 field.
 
 In their seminal paper Edwards and Anderson$^{6}$ (EA) introduced a lattice
 model for the spin glass, with short-range interactions, and used a replica
 method to perform the average over quenched disorder. Two different pictures
 have since then been proposed for the spin glass. One is the mean-field
 theory for the spin glasses, provided by the Parisi$^{7}$ solution for the
 infinite-range, Sherrington-Kirkpatrick (SK) model,$^{8}$ which predicts a
 glassy phase described by an infinite number of pure states organized in an
 ultrametric structure, and a phase transition occurring in a magnetic field.
 The alternative is the ''droplet'' model,$^{9-11}$ which claims that the
 real, short-range spin systems behave quite differently, the glassy phase
 being described by only two pure states, related by a global inversion of
 the spins, and no phase transition occurring in a magnetic field. The first
 picture results from replica symmetry breaking, while in the second picture
 there is no replica symmetry breaking. The spin glass transition in a field,
 found by de Almeida and Thouless in the SK model,$^{12}$ occurs along a line
 in the field-temperature plane, the AT-line, which marks the instability of
 the replica-symmetric solution against replica symmetry breaking. In the
 mean-field theory, it represents a line of second-order transitions. Via the
 replica method the order parameter is represented by a field $Q_{\alpha
 \beta }$, (where $\alpha =1,\ldots ,n$ is a replica label), which has 
 $n(n-1)/2$ independent components (since 
 $Q_{\alpha \alpha }\equiv 0$, $Q_{\alpha \beta }\equiv Q_{\beta \alpha }$). 
 Linear combinations of these
 components define three different sets of modes: the replicon, the anomalous
 and the longitudinal.$^{13}$ In zero-field all the modes become critical at
 the transition temperature, while in a finite-field only the replicon modes
 become critical at the AT-line.

 The existence, or not, of a spin glass transition in a field, is a crucial
 issue in the characterization of the spin glass. It has been disputed both
 from the theoretical as well as the experimental side.$^{14,15}$ Extensive
 numerical simulations$^{16}$ have also been carried out to clarify the
 situation, but have neither been conclusive. A fundamental step towards the
 clarification of the controversy, and therefore the understanding of the
 spin glass, lies in the investigation of how the fluctuations, associated to
 the finite-range interactions, will modify the mean-field picture, and in
 particular the AT-transition. Green, Moore and Bray$^{17}$ made a one-loop
 perturbation calculation of the AT-line in a magnetic field, however
 considering the effect of the field only in the masses, the couplings
 remaining the same as in zero-field. It becomes then most important to study
 the problem of the spin glass in a field by a powerful analytical method,
 such as the renormalization group (RG), the aim being to find a fixed-point
 that controls the spin glass transition in a non-zero field. The spin glass
 transition in zero-field was already studied within the RG by Harris 
 \textit{et} \textit{al.}$^{18}$ Bray and Roberts$^{19}$ considered the case of
 non-zero field and carried out a RG study, in which they retained only the
 replicon modes to calculate the critical behavior at the AT-line.
 They found no physical fixed-point in a field.

 In this paper we study the spin glass transition in a field, presenting the
 complete set of RG equations for an Ising spin glass in a uniform magnetic
 field, which contains the replicon, anomalous and longitudinal modes, and
 explicit dependence on the number of replicas $n$. This allow us to discuss,
 in a common framework, the transitions in zero and non-zero field, and the
 crossover region around the zero-field critical point, thus investigating
 the role that a small magnetic field plays in the transition. We keep the
 replica number $n$ finite to discuss the limit $n\rightarrow 0$ in the
 calculation, which is a delicate and important matter when the anomalous and
 longitudinal modes are considered. Here we analyze the case where the limit 
 $n\rightarrow 0$ corresponds to a theory with a symmetry that involves
 equality between the anomalous and longitudinal masses, as it occurs in the
 spin glass mean field theory. By contrast, the case where there may be
 singularities in the limit $n\rightarrow 0$, which may lead to the breaking
 up of the equality between the anomalous and longitudinal masses, will be
 considered in a separate publication.$^{20}$ The main purpose of our work is
 then to search for the possible existence of a fixed-point associated to the
 spin glass transition in a field, within the complete set of RG equations.
 To avoid the complexity of the glassy phase, we approach the transition from
 the high-temperature, replica-symmetric phase, like Bray and Roberts.$^{19}$

 In a previous publication,$^{21}$ we presented the derivation of the replica
 field theory, relevant to the study of the spin glass transition of
 short-range models in a field. We argued that for studying this transition
 around the upper critical dimension $d=6$, it is necessary to use the
 generic theory, with all the three bare masses and the eight cubic couplings
 involving the replica fields, which correspond to all the possible
 invariants of the replica symmetric Lagrangian. It is however more
 convenient to work directly with a field theory defined in terms of the
 eigenfields, which corresponds to a block-diagonalization of the mass
 operator into the replicon, anomalous and longitudinal modes, and we
 introduced then the appropriate representation of the cubic interaction. The
 mass operator and the cubic couplings for the eigenfields, are in fact
 simply obtained by performing a formal decomposition of the replica fields
 into the replicon, anomalous and longitudinal fields, and use the generic
 properties of the fields. However, in order to do a perturbation calculation
 one still has to deal with the problem of having non-diagonal free
 propagators. Such a difficuly was overcome in Ref. 21, by introducing a
 particular non-orthogonal basis and its biorthogonal counterpart, in which a
 set of vertices were computed, the perturbation computation being alliviated
 through selection rules. In this work we present a different approach, in
 which we introduce projection operators into the non-diagonal propagators,
 the projectors being simply derived from the generic properties of the
 fields. This method presents a generalization, for the complete set of
 modes, of the method used by Bray and Roberts.$^{19}$ We have then a field
 theory for the spin glass that works directly with the longitudinal,
 anomalous and replicon fields, and allows to perform a standard perturbation
 expansion. An alternative formulation of the spin glass field theory can
  be obtained using Replica Fourier Transform methods, which turn out 
 particularly efficient in the study of the replica symmetry 
 broken phase.$^{22}$

 The outline of the paper is as follows. In Sec. II we present the cubic
 field theory written in terms of the replicon, anomalous and longitudinal
 fields. In Sec. III we derive the mean-field results. In Sec. IV we present
 the renormalization group equations, calculated to lowest order in 
 $\varepsilon =6-d$ (where $d$ is the spatial dimension), and discuss the spin
 glass transitions in zero and non-zero field. Sec.V contains the conclusions
 of our work.

 \section{CUBIC FIELD THEORY}

 We consider a short-range Ising spin glass in a uniform magnetic field $H$,
 described by the Edwards-Anderson model,

 \begin{equation}
 \mathcal{H}=-{\sum_{(ij)} }J_{ij}S_{i}S_{j}-H{\sum_{i} }
 S_{i}  \label{1}
 \end{equation}

 \noindent for $N$ spins, $S_{i}=\pm 1$, located on a regular $d$-dimensional
 lattice, where the bonds $J_{ij}$, which couple nearest-neighbor spins only,
 are independent random variables with a Gaussian distribution, characterized
 by zero mean and variance $\Delta ^{2}$. The summations are over pairs $(ij)$
 of distinct sites on the lattice and over the lattice sites $i$. The replica
 method allows to calculate the average of the free energy over the quenched
 disorder, in terms of the average of $n$ replicas of the partition function $
 \overline{Z^{n}}$, with $n$ positive integer, which provides the spin glass
 behavior in the analytically continued limit $n\rightarrow 0$. This
 procedure transforms the originally disordered system into a uniform one,
 described by an effective Hamiltonian for the replica spins $S_{i}^{\alpha }$
 , $\alpha =1,\ldots ,n$. A field theoretical continuum representation of the
 spin glass lattice model can be built using a standard Hubbard-Stratonovich
 transformation, which leads to the averaged replicated partition function,
 expressed as an integral over replica fields $Q_{i}^{\alpha \beta }$.$^{21}$
 The fields $Q_{i}^{\alpha \beta }$ are defined on a $n(n-1)/2$-dimensional
  replica space, of the pairs $(\alpha \beta )$ of distinct
 replicas.
 
 In order to construct a perturbation expansion around the mean-field
 solution, which corresponds to the infinite range or infinite dimensional
 (i.e. spin coordination number $z\rightarrow \infty $) model, one separates
 the field $Q_{i}^{\alpha \beta }$ into
 
 \begin{equation}
 Q_{i}^{\alpha \beta }=Q^{\alpha \beta }+\phi _{i}^{\alpha \beta }  \label{2}
 \end{equation}
 where $Q^{\alpha \beta }$ represents the uniform, mean-field value of the
 order parameter, and $\phi _{i}^{\alpha \beta }$ are the fluctuations around
 it\textit{. }In addition, since we approach the transition from the
 high-temperature phase, one can assume a replica symmetric mean-field
 solution
 
 \begin{equation}
 Q^{\alpha \beta }=Q.  \label{3}
 \end{equation}
 
 \noindent The partition function then takes the form
 
 \begin{equation}
 \overline{Z^{n}}\sim \int \mathcal{D}\phi \exp \left( -\mathcal{L}^{(1)}-
 \mathcal{L}^{(2)}-\mathcal{L}^{(3)}-\ldots \right)  \label{4}
 \end{equation}
 
 \noindent where, after Fourier transform into momenta space, one has, for
 the contributions up to cubic order,
 
 \begin{eqnarray}
 \mathcal{L}^{(1)} &=&\sqrt{N}{\sum_{(\alpha \beta )} }\left(
 Q\Theta ^{-1}-\ll S^{\alpha }S^{\beta }\gg \right) \phi _{\mathbf{p}
 =0}^{\alpha \beta }  \label{5} \\
 \mathcal{L}^{(2)} &=&\frac{1}{2}{\sum_{(\alpha \beta )(\gamma \delta )} }
 {\sum_{\mathbf{p}} }\phi _{\mathbf{p}}^{\alpha \beta
 }M_{\alpha \beta ,\gamma \delta }(\mathbf{p})\phi _{-\mathbf{p}}^{\gamma
 \delta }  \label{6} \\
 \mathcal{L}^{(3)} &=&-\frac{1}{3!}\frac{1}{\sqrt{N}}
 {\sum_{(\alpha\beta )(\gamma \delta )(\mu \nu )} }
 {\sum_{\mathbf{p}_{1},\mathbf{p}_{2},\mathbf{p}_{3}} }^{\prime }
 W_{\alpha \beta ,\gamma \delta ,\mu
 \nu }\phi _{\mathbf{p}_{1}}^{\alpha \beta }\phi _{\mathbf{p}_{2}}^{\gamma
 \delta }\phi _{\mathbf{p}_{3}}^{\mu \nu }  \label{7}
 \end{eqnarray}
 the prime in the sum in Eq. (7) indicating constraint to momentum
 conservation, $\mathbf{p}_{1}+\mathbf{p}_{2}+\mathbf{p}_{3}=0$. The quantity 
 $\Theta ^{-1}=(k_{B}T/\Delta )^{2}/z$, where the coordination number is 
 $z=2d $, essentially represents the temperature squared. The average 
 $\ll\ldots \gg $ is defined with the Boltzmann weight, $\exp 
 \left( {\sum_{(\alpha \beta )} }
 QS^{\alpha }S^{\beta }+h{\sum_{\alpha } }
 S^{\alpha }\right) $, where $h=H/k_{B}T$. The masses $M_{\alpha \beta
 ,\gamma \delta }$ and the couplings $W_{\alpha \beta ,\gamma \delta ,\mu \nu
 }$ are expressed in terms of spin correlations. The sum in the first
 Brillouin zone are confined to the range $0<\left| \mathbf{p}\right|
 <\Lambda $, with cutoff $\Lambda \simeq 1$, and $M_{\alpha \beta ,\gamma
 \delta }(\mathbf{p})$ is expanded for $\mathbf{p}\ll \mathbf{1}$, keeping as
 usual only the terms up to second order. Then, an appropriate rescaling of
 the fields $(\phi (z\Theta )^{-1/2}\rightarrow \phi )$, with a corresponding
 rescaling of the masses and the couplings, allows to write the mass operator
 in a standard form, i.e. with the coefficient of the momentum equal to unity,
 
 \begin{eqnarray}
 M_{\alpha \beta ,\gamma \delta }(\mathbf{p}) &=&\mathbf{p}^{2}\delta
 _{\alpha \beta ,\gamma \delta }^{Kr}+z\left[ \delta _{\alpha \beta ,\gamma
 \delta }^{Kr}\right. -  \nonumber \\
 &&-\left. \Theta \left( \ll S^{\alpha }S^{\beta }S^{\gamma }S^{\delta }\gg
 -\ll S^{\alpha }S^{\beta }\gg \ll S^{\gamma }S^{\delta }\gg \right) \right] ,
 \label{8}
 \end{eqnarray}
 the Kronecker-delta being defined in the space of replica pairs. For the
 interaction operator one has,
 
 \begin{eqnarray}
 W_{\alpha \beta ,\gamma \delta ,\mu \nu } &=&(z\Theta )^{3/2}\left[ \ll
 S^{\alpha }S^{\beta }S^{\gamma }S^{\delta }S^{\mu }S^{\nu }\gg \right. -\ll
 S^{\alpha }S^{\beta }\gg \ll S^{\gamma }S^{\delta }S^{\mu }S^{\nu }\gg 
 \nonumber \\
 &-&\ll S^{\gamma }S^{\delta }\gg \ll S^{\alpha }S^{\beta }S^{\mu }S^{\nu
 }\gg -\ll S^{\mu }S^{\nu }\gg \ll S^{\alpha }S^{\beta }S^{\gamma }S^{\delta
 }\gg  \nonumber \\
 &&+\left. 2\ll S^{\alpha }S^{\beta }\gg \ll S^{\gamma }S^{\delta }\gg \ll
 S^{\mu }S^{\nu }\gg \right] .  \label{9}
 \end{eqnarray}
 
 \noindent The expansion in the fluctuations corresponds to an expansion in 
 $1/z$.
 
 Replica symmetry allows three distinct components for the mass, Eq. (8), and
 eight distinct components for the cubic interaction, Eq. (9). Hence,
 
 \begin{eqnarray}
 \mathcal{L}^{(2)} &=&\frac{1}{2}{\sum_{\mathbf{p}} }\left\{ (
 \mathbf{p}^{2}\mathbf{+}M_{1}){\sum_{(\alpha \beta )} }\phi _{
 \mathbf{p}}^{\alpha \beta }\phi _{-\mathbf{p}}^{\alpha \beta }\right. +M_{2}
 {\sum_{(\alpha \beta \gamma )} }(\phi _{\mathbf{p}}^{\alpha \beta
 }\phi _{-\mathbf{p}}^{\alpha \gamma }+\phi _{\mathbf{p}}^{\alpha \beta }\phi
 _{-\mathbf{p}}^{\beta \gamma })  \nonumber \\
 &&\qquad \qquad +\left. M_{3}{\sum_{(\alpha \beta \gamma \delta )}}
 \phi _{\mathbf{p}}^{\alpha \beta }\phi _{-\mathbf{p}}^{\gamma \delta
 }\right\}  \label{10}
 \end{eqnarray}
 
 \noindent and
 
 \begin{eqnarray}
 \mathcal{L}^{(3)} &=&-\frac{1}{\sqrt{N}}
 {\sum_{\mathbf{p}_{1},\mathbf{p}_{2},\mathbf{p}_{3}} }^{\prime }
 \left\{ \frac{1}{6}W_{1}
 {\sum_{(\alpha ,\beta ,\gamma )} }\phi _{\mathbf{p}_{1}}^{\alpha \beta }
 \phi _{
 \mathbf{p}_{2}}^{\beta \gamma }\phi _{\mathbf{p}_{3}}^{\gamma \alpha
 }\right. +\frac{1}{12}W_{2}{\sum_{(\alpha \beta )} }\phi _{\mathbf{
 p}_{1}}^{\alpha \beta }\phi _{\mathbf{p}_{2}}^{\alpha \beta }\phi _{\mathbf{p
 }_{3}}^{\alpha \beta }  \nonumber \\
 &&+\frac{1}{2}W_{3}{\sum_{(\alpha ,\beta ,\gamma )} }\phi _{
 \mathbf{p}_{1}}^{\alpha \beta }\phi _{\mathbf{p}_{2}}^{\alpha \beta }\phi _{
 \mathbf{p}_{3}}^{\beta \gamma }+\frac{1}{8}W_{4}
 {\sum_{(\alpha ,\beta,\gamma ,\delta )} }
 \phi _{\mathbf{p}_{1}}^{\alpha \beta }\phi _{
 \mathbf{p}_{2}}^{\alpha \beta }\phi _{\mathbf{p}_{3}}^{\gamma \delta } 
 \nonumber \\
 &&+\frac{1}{2}W_{5}{\sum_{(\alpha ,\beta ,\gamma ,\delta )} }\phi
 _{\mathbf{p}_{1}}^{\alpha \beta }\phi _{\mathbf{p}_{2}}^{\beta \gamma }\phi
 _{\mathbf{p}_{3}}^{\gamma \delta }+\frac{1}{6}W_{6}
 {\sum_{(\alpha,\beta ,\gamma ,\delta )} }
 \phi _{\mathbf{p}_{1}}^{\alpha \beta }\phi _{
 \mathbf{p}_{2}}^{\alpha \gamma }\phi _{\mathbf{p}_{3}}^{\alpha \delta }
 \label{11} \\
 &&+\frac{1}{4}W_{7}{\sum_{(\alpha ,\beta ,\gamma ,\delta ,\mu )} }
 \phi _{\mathbf{p}_{1}}^{\alpha \beta }\phi _{\mathbf{p}_{2}}^{\alpha \gamma
 }\phi _{\mathbf{p}_{3}}^{\delta \mu }+\left. \frac{1}{48}W_{8}
 {\sum_{(\alpha ,\beta ,\gamma ,\delta ,\mu ,\nu )} }
 \phi _{\mathbf{p}_{1}}^{\alpha \beta }\phi _{\mathbf{p}_{2}}^{\gamma \delta }\phi _{\mathbf{p}
 _{3}}^{\mu \nu }\right\} ,  \nonumber
 \end{eqnarray}
 
 \noindent the sums in Eqs. (10)-(11) being restricted to distinct replicas.
 One can rewrite $\mathcal{L}^{(2)}$ and $\mathcal{L}^{(3)}$ in terms of sums
 over unrestricted replicas, for which we obtain forms similar to those, with
 new masses and new couplings, defined as linear combinations of $M_{i}$ and 
 $W_{i}$.$^{21}$ From a symmetry point of view, Eqs. (10)-(11) contain all the
 possible quadratic and cubic invariants of the symmetry group of the system,
 which corresponds to invariance of the Lagrangian under any permutation of
 the $n$ replicas.
 
 We now wish to write the quadratic part $\mathcal{L}^{(2)}$ in terms of the
 mass eigenvalues, which corresponds to a block diagonalization into the
 longitudinal (L), anomalous (A) and replicon (R) subspaces. Any replica
 field $\phi ^{\alpha \beta }$ can be decomposed in its projections onto the
 L, A, R subspaces,
 
 \begin{equation}
 \phi ^{\alpha \beta }=\phi _{\alpha \beta }^{L}+\phi _{\alpha \beta
 }^{A}+\phi _{\alpha \beta }^{R}.  \label{12}
 \end{equation}
 The field projections $\phi _{\alpha \beta }^{i}$ can be obtained via the
 projection operators $P_{\alpha \beta ,\gamma \delta }^{i}$, onto the
 different subspaces, 
 \begin{equation}
 \phi _{\alpha \beta }^{i}={\sum_{(\gamma \delta )} }P_{\alpha
 \beta ,\gamma \delta }^{i}\phi ^{\gamma \delta },\quad i=L,A,R.  \label{13}
 \end{equation}
 The matrix elements of the projection operators are simply derived from the
 general properties of the replicon, anomalous and longitudinal fields, as
 shown in the Appendix. The three subspaces are characterized by different
 symmetries, as follows.
 
 The longitudinal subspace is $1-$dimensional. The longitudinal eigenvector 
 $\mathbf{e}^{L}$\ is symmetric under interchange of all replica indices. This
 implies a replica independent longitudinal field
 
 \begin{equation}
 \phi _{\alpha \beta }^{L}=\phi ^{L}  \label{14}
 \end{equation}
 given by
 
 \begin{equation}
 \phi ^{L}(\mathbf{p})=a^{L}(\mathbf{p})e^{L}  \label{15}
 \end{equation}
 and a replica independent longitudinal projector
 
 \begin{equation}
 P_{\alpha \beta ,\gamma \delta }^{L}=P^{L}  \label{16}
 \end{equation}
 defined by
 
 \begin{equation}
 P^{L}=e^{L}e^{L},
 \end{equation}
 
 \noindent with $e^{L}$\ in Eqs. (15) and (17), representing a component of
 the replica independent eigenvector, $e_{\alpha \beta }^{L}=e^{L}$. One
 finds for the longitudinal projector the explicit form, Eq. (A3),
 
 \begin{equation}
 P^{L}=\frac{2}{n(n-1)}.  \label{18}
 \end{equation}
 The longitudinal subspace is hence described by a single scalar field 
 $\phi_{L}$.
 
 The anomalous subspace is $(n-1)-$dimensional. The anomalous eigenvectors
 are symmetric under interchange of all but one of the replica indices. This
 implies that a generic anomalous field can be represented by a one-replica
 field $\phi _{\alpha }^{A}$, i.e. it can be expressed as,
 
 \begin{equation}
 \phi _{\alpha \beta }^{A}=\frac{1}{2}(\phi _{\alpha }^{A}+\phi _{\beta
 }^{A}),  \label{19}
 \end{equation}
 with the condition
 
 \begin{equation}
 {\sum_{\alpha } }\phi _{\alpha }^{A}=0.  \label{20}
 \end{equation}
 One can write the one-replica field $\phi _{\alpha }^{A}$\ in terms of the
 set of anomalous eigenvectors $\mathbf{e}^{A,\mu }$, as 
 \begin{equation}
 \phi _{\alpha }^{A}(\mathbf{p})={\sum_{\mu =1}^{n-1}}a^{A,\mu }(
 \mathbf{p})e_{\alpha }^{A,\mu },  \label{21}
 \end{equation}
 with the normalization ${\sum_{\alpha } }e_{\alpha
 }^{A,\mu }e_{\alpha }^{A,\mu ^{\prime }}=\frac{4}{(n-2)}\delta _{\mu \mu
 ^{\prime }}$. One then defines an anomalous projector operator
 
 \begin{equation}
 P_{\alpha ,\beta }^{A}={\sum_{\mu =1}^{n-1}}e_{\alpha }^{A,\mu
 }e_{\beta }^{A,\mu },  \label{22}
 \end{equation}
 which has the property
 
 \begin{equation}
 {\sum_{\alpha } }P_{\alpha ,\beta }^{A}=0.  \label{23}
 \end{equation}
 The projectors in Eqs. (13) and (22) are simply related by
 
 \begin{equation}
 P_{\alpha \beta ,\gamma \delta }^{A}=\frac{1}{4}\left[ P_{\alpha ,\gamma
 }^{A}+P_{\alpha ,\delta }^{A}+P_{\beta ,\gamma }^{A}+P_{\beta ,\delta
 }^{A}\right] .  \label{24}
 \end{equation}
 The explicit form of the anomalous projector $P_{\alpha ,\beta }^{A}$\ is
 given by, Eq. (A8),
 
 \begin{equation}
 P_{\alpha ,\beta }^{A}=\frac{4}{(n-2)}\left( \delta _{\alpha \beta }-\frac{1
 }{n}\right) .  \label{25}
 \end{equation}
 The anomalous subspace is then described by the set of fields $\phi _{\alpha
 }^{A}$, $\alpha =1,\ldots ,n$, with the constraint Eq. (20), which leaves 
 $(n-1)$\ independent fields.
 
 The replicon subspace is $n(n-3)/2-$dimensional. The replicon
 eigenvectors are symmetric under interchange of all but two replica indices.
 This implies that the replicon fields depend on two replica indices and
 verify the conditions$^{23}$
 
 \begin{equation}
 {\sum_{\alpha (\neq \beta )} }\phi _{\alpha \beta }^{R}=0,\quad
 \alpha =1,\ldots ,n.  \label{26}
 \end{equation}
 A replicon field can be written in terms of the set of replicon orthonormal
 eigenvectors $\mathbf{e}^{R,\nu}$as
 
 \begin{equation}
 \phi _{\alpha \beta }^{R}(\mathbf{p})={\sum_{\nu =1}^{\frac{1}{2}
 n(n-3)}}a^{R,\nu }(\mathbf{p})e_{\alpha \beta }^{R,\nu }.  \label{27}
 \end{equation}
 The replicon projector is defined by
 
 \begin{equation}
 P_{\alpha \beta ,\gamma \delta }^{R}={\sum_{\nu =1}^{\frac{1}{2}
 n(n-3)}}e_{\alpha \beta }^{R,\nu }e_{\gamma \delta }^{R,\nu }  \label{28}
 \end{equation}
 and has the property
 
 \begin{equation}
 {\sum_{\alpha (\neq \beta )} }P_{\alpha \beta ,\gamma \delta
 }^{R}=0.  \label{29}
 \end{equation}
 
 \noindent The explicit form of the replicon projector is given by, Eq.
 (A14),
 
 \begin{eqnarray}
 P_{\alpha \beta ,\gamma \delta }^{R} &=&\left( \delta _{\alpha \gamma
 }\delta _{\beta \delta }+\delta _{\alpha \delta }\delta _{\beta \gamma
 }\right) -\left( \delta _{\alpha \gamma }+\delta _{\alpha \delta }+\delta
 _{\beta \gamma }+\delta _{\beta \delta }\right) \frac{1}{n-2}  \nonumber \\
 &&+\frac{2}{(n-1)(n-2)}  \label{30}
 \end{eqnarray}
 
 \noindent naturally, with $\alpha \neq \beta $ and $\gamma \neq \delta $.
 Eq. (30) is equivalent to the replicon projector as introduced by Bray and
 Roberts.$^{19}$ The replicon subspace is then described by the set of $\phi
 _{\alpha \beta }^{R}$ fields, with the constraints in Eq. (26), which leaves 
 $n(n-1)/2-n=n(n-3)/2$ independent fields. The characteristic property of
 projection operators $\mathbf{P}^{2}=\mathbf{P}$ is readily observed for the
 longitudinal, anomalous and replicon projectors in Eqs. (18), (25) and (30).
 
 Introducing the field decomposition in Eq. (12), with the definitions in
 Eqs. (14) and (19), into Eq. (10), and using repeatedly the conditions in
 Eqs. (20) and (26), one obtains the quadratic part in the diagonalized form 
 \begin{eqnarray}
 \medskip \mathcal{L}^{(2)} &=&\frac{1}{2}{\sum_{\mathbf{p}} }
 \left\{ \left( \mathbf{p}^{2}+m_{L}\right) \frac{n(n-1)}{2}\phi ^{L}(\mathbf{
 p})\phi ^{L}(-\mathbf{p})\right.  \nonumber \\
 &&+\left( \mathbf{p}^{2}+m_{A}\right) \frac{(n-2)}{4}
 {\sum_{\alpha } }\phi _{\alpha }^{A}(\mathbf{p})\phi _{\alpha }^{A}
 (-\mathbf{p})
 \label{31} \\
 &&+\left. \left( \mathbf{p}^{2}+m_{R}\right) 
 {\sum_{(\alpha \beta )} }\phi _{\alpha \beta }^{R}(\mathbf{p})
 \phi _{\alpha \beta }^{R}(-
 \mathbf{p})\right\}  \nonumber
 \end{eqnarray}
 \noindent where $m_{L}$, $m_{A}$, $m_{R}$ are the longitudinal, anomalous
 and replicon mass eigenvalues, which are given by
 
 \begin{eqnarray}
 m_{L} &=&M_{1}+2(n-2)M_{2}+\frac{1}{2}(n-2)(n-3)M_{3}.  \nonumber \\
 m_{A} &=&M_{1}+(n-4)M_{2}-(n-3)M_{3}  \label{32} \\
 m_{R} &=&M_{1}-2M_{2}+M_{3}.  \nonumber
 \end{eqnarray}
 
 \noindent We notice that the anomalous and longitudinal masses are equal in
 the limit $n\rightarrow 0$. For later discussion we define the variable
 
 \begin{equation}
 \bar{m}_{AL}=\frac{m_{A}-m_{L}}{n},  \label{33}
 \end{equation}
 
 \noindent which is given by
 
 \begin{equation}
 \bar{m}_{AL}=-M_{2}-\frac{1}{2}(n-3)M_{3}.  \label{34}
 \end{equation}
 
 Now, because of Eqs. (20) and (26), the fields $\phi _{\alpha }^{A}$ and 
 $\phi _{\alpha \beta }^{R}$ are still not independent. One can however
 construct an expansion in terms of the orthonormal eigenvectors 
 $\mathbf{e}^{L}$, $\mathbf{e}^{A,\mu }$, $\mathbf{e}^{R,\nu }$. 
 Introducing Eqs. (15), (21) and (27) into
 Eq. (31), one gets
 
 \begin{eqnarray}
 \medskip \mathcal{L}^{(2)} &=&\frac{1}{2}{\sum_{\mathbf{p}} }
 \left\{ \left( \mathbf{p}^{2}+m_{L}\right) a^{L}(\mathbf{p})a^{L}(-\mathbf{p}
 )\right.  \nonumber \\
 &&+\left( \mathbf{p}^{2}+m_{A}\right) {\sum_{\mu =1}^{n-1}}
 a^{A,\mu }(\mathbf{p})a^{A,\mu }(-\mathbf{p})  \label{35} \\
 &&+\left. \left( \mathbf{p}^{2}+m_{R}\right){\sum_{\nu =1}^{\frac{1}{2}
 n(n-3)}}a^{R,\nu }(\mathbf{p})a^{R,\nu }(-\mathbf{p})\right\}  \nonumber
 \end{eqnarray}
 
 \noindent and then
 
 \begin{eqnarray}
 &<&a^{L}(\mathbf{p})a^{L}(-\mathbf{p})>=\left( \mathbf{p}^{2}+m_{L}\right)
 ^{-1}  \nonumber \\
 &<&a^{A,\mu }(\mathbf{p})a^{A,\mu ^{\prime }}(-\mathbf{p})>=\left( \mathbf{p}
 ^{2}+m_{A}\right) ^{-1}\delta _{\mu ,\mu ^{\prime }}  \label{36} \\
 &<&a^{R,\nu }(\mathbf{p})a^{A,\nu ^{\prime }}(-\mathbf{p})>=\left( \mathbf{p}
 ^{2}+m_{R}\right) ^{-1}\delta _{\nu ,\nu ^{\prime }}.  \nonumber
 \end{eqnarray}
 
 \noindent Using the expressions for the fields in Eqs. (15), (21), (27),
 together with Eq. (36), and the definitions in Eqs. (17), (22), (28), one
 finds the bare propagators
 
 \begin{eqnarray}
 &<&\phi ^{L}(\mathbf{p})\phi ^{L}(-\mathbf{p})>=\left( \mathbf{p}
 ^{2}+m_{L}\right) ^{-1}P^{L}  \nonumber \\
 &<&\phi _{\alpha }^{A}(\mathbf{p})\phi _{\beta }^{A}(-\mathbf{p})>=\left( 
 \mathbf{p}^{2}+m_{A}\right) ^{-1}P_{\alpha ,\beta }^{A}  \label{37} \\
 &<&\phi _{\alpha \beta }^{R}(\mathbf{p})\phi _{\gamma \delta }^{R}(-\mathbf{p
 })>=\left( \mathbf{p}^{2}+m_{R}\right) ^{-1}P_{\alpha \beta ,\gamma \delta
 }^{R}.  \nonumber
 \end{eqnarray}
 
 The cubic interaction can be written in terms of the longitudinal, anomalous
 and replicon fields, by following a procedure similar to the one that lead
 to Eq. (31), that is, introducing the decomposition in Eq. (12), with Eqs.
 (14) and (19), into Eq. (11) and repeatedly applying the conditions in Eqs.
 (20) and (26), one obtains
 
 \begin{eqnarray}
 \mathcal{L}^{(3)} &=&-\frac{1}{\sqrt{N}}
 {\sum_{\mathbf{p}_{1},\mathbf{p}_{2},\mathbf{p}_{3}} }^{\prime }
 \left\{ \frac{1}{6}g_{1}\right. 
 {\sum_{\alpha ,\beta ,\gamma } }\phi _{\alpha \beta }^{R}(\mathbf{p
 }_{1})\phi _{\beta \gamma }^{R}(\mathbf{p}_{2})\phi _{\gamma \alpha }^{R}(
 \mathbf{p}_{3})  \nonumber \\
 &&+\frac{1}{12}g_{2}{\sum_{\alpha , \beta } }\phi _{\alpha \beta
 }^{R}(\mathbf{p}_{1})\phi _{\alpha \beta }^{R}(\mathbf{p}_{2})\phi _{\alpha
 \beta }^{R}(\mathbf{p}_{3})  \label{38} \\
 &&+\frac{1}{4}g_{3}{\sum_{\alpha , \beta } }\phi _{\alpha \beta
 }^{R}(\mathbf{p}_{1})\phi _{\alpha \beta }^{R}(\mathbf{p}_{2})\left( \phi
 _{\alpha }^{A}(\mathbf{p}_{3})+\phi _{\beta }^{A}(\mathbf{p}_{3})\right) 
 \nonumber \\
 &&+\frac{1}{2}g_{4}{\sum_{\alpha , \beta } }\phi _{\alpha \beta
 }^{R}(\mathbf{p}_{1})\phi _{\alpha \beta }^{R}(\mathbf{p}_{2})\phi ^{L}(
 \mathbf{p}_{3})+\frac{1}{2}g_{5}{\sum_{\alpha , \beta } }\phi
 _{\alpha \beta }^{R}(\mathbf{p}_{1})\phi _{\alpha }^{A}(\mathbf{p}_{2})\phi
 _{\beta }^{A}(\mathbf{p}_{3})  \nonumber \\
 &&+\frac{1}{6}g_{6}{\sum_{\alpha } }\phi _{\alpha }^{A}(\mathbf{p}
 _{1})\phi _{\alpha }^{A}(\mathbf{p}_{2})\phi _{\alpha }^{A}(\mathbf{p}_{3})+
 \frac{1}{2}g_{7}{\sum_{\alpha } }\phi _{\alpha }^{A}(\mathbf{p}
 _{1})\phi _{\alpha }^{A}(\mathbf{p}_{2})\phi ^{L}(\mathbf{p}_{3})  \nonumber
 \\
 &&+\left. \frac{1}{6}g_{8}\phi ^{L}(\mathbf{p}_{1})\phi ^{L}(\mathbf{p}
 _{2})\phi ^{L}(\mathbf{p}_{3})\right\}  \nonumber
 \end{eqnarray}
 
 \noindent with the couplings,
 
 \begin{eqnarray}
 g_{1} &=&W_{1}-3W_{5}+3W_{7}-W_{8}  \nonumber \\
 g_{2} &=&W_{2}-6W_{3}+3W_{4}+6W_{5}+4W_{6}-12W_{7}+4W_{8}  \label{39} \\
 g_{3} &=&-W_{1}+\frac{1}{2}W_{2}+\frac{1}{2}\left( n-8\right) W_{3}-\frac{1}{
 2}\left( n-5\right) W_{4}-\left( n-8\right) W_{5}  \nonumber \\
 &&-\frac{1}{2}\left( n-6\right) W_{6}+\frac{1}{2}\left( 5n-28\right)
 W_{7}-\left( n-5\right) W_{8}  \nonumber \\
 g_{4} &=&-W_{1}+\frac{1}{2}W_{2}+\left( n-4\right) W_{3}+\frac{1}{4}\left(
 n^{2}-5n+10\right) W_{4}-2\left( n-4\right) W_{5}  \nonumber \\
 &&-\left( n-3\right) W_{6}-\frac{1}{2}\left( n-4\right) \left( n-7\right)
 W_{7}+\frac{1}{4}\left( n-4\right) \left( n-5\right) W_{8}  \nonumber \\
 g_{5} &=&\frac{1}{4}(n-4)W_{1}+\frac{1}{4}W_{2}+\frac{1}{2}\left( n-5\right)
 W_{3}-\frac{1}{4}\left( 2n-7\right) W_{4}  \nonumber \\
 &&+\frac{1}{4}\left( n-5\right) \left( n-6\right) W_{5}-\frac{1}{2}\left(
 n-4\right) W_{6}-\frac{1}{4}\left( n-4\right) \left( 2n-13\right) W_{7} 
 \nonumber \\
 &&+\frac{1}{4}\left( n-4\right) \left( n-5\right) W_{8}  \nonumber \\
 g_{6} &=&-\frac{1}{4}(3n-8)W_{1}+\frac{1}{8}(n-4)W_{2}+\frac{3}{8}\left(
 n-4\right) ^{2}W_{3}  \nonumber \\
 &&-\frac{3}{8}\left( n-3\right) \left( n-4\right) W_{4}-\frac{3}{2}\left(
 n-3\right) \left( n-4\right) W_{5}  \nonumber \\
 &&+\frac{1}{8}\left( n-3\right) \left( n^{2}-6n+16\right) W_{6}-\frac{3}{8}
 \left( n-3\right) \left( n-4\right) \left( n-8\right) W_{7}  \nonumber \\
 &&+\frac{1}{4}\left( n-3\right) \left( n-4\right) \left( n-5\right) W_{8} 
 \nonumber \\
 g_{7} &=&\left( n-2\right) \left[ \frac{1}{4}\left( n-4\right) W_{1}\right. +
 \frac{1}{4}W_{2}+\left( n-3\right) W_{3}+\frac{1}{8}\left( n-3\right) \left(
 n-6\right) W_{4}  \nonumber \\
 &&+\frac{1}{2}\left( n-3\right) \left( n-6\right) W_{5}+\frac{1}{4}\left(
 n-3\right) \left( n-4\right) W_{6}  \nonumber \\
 &&+\frac{1}{8}\left( n-3\right) \left( n-4\right) \left( n-12\right)
 W_{7}-\left. \frac{1}{8}\left( n-3\right) \left( n-4\right) \left(
 n-5\right) W_{8}\right]  \nonumber \\
 g_{8} &=&n(n-1)\left[ \left( n-2\right) W_{1}+\frac{1}{2}W_{2}\right.
 +3\left( n-2\right) W_{3}  \nonumber \\
 &&+\frac{3}{4}\left( n-2\right) \left( n-3\right) W_{4}+3\left( n-2\right)
 \left( n-3\right) W_{5}  \nonumber \\
 &&+\left( n-2\right) \left( n-3\right) W_{6}+\frac{3}{2}\left( n-2\right)
 \left( n-3\right) \left( n-4\right) W_{7}  \nonumber \\
 &&+\left. \frac{1}{8}\left( n-2\right) \left( n-3\right) \left( n-4\right)
 \left( n-5\right) W_{8}\right] .  \nonumber
 \end{eqnarray}
 
 \noindent Again, we notice that the couplings $g_{3}$ and $g_{4}$, $g_{6}$
 and $g_{7}$, $g_{7}$ and $g_{8}/n$, are equal in the limit $n\rightarrow 0$.
 For later discussion we define the variables 
 \begin{eqnarray}
 \bar{g}_{4} &=&\frac{4}{n}(g_{4}-g_{3})  \nonumber \\
 \bar{g}_{7} &=&\frac{2}{n}(g_{7}-g_{6})  \label{40} \\
 \bar{g}_{8} &=&\frac{1}{n^{3}}(g_{8}-3ng_{7}+2ng_{6})  \nonumber
 \end{eqnarray}
 
 \noindent for which we have
 
 \begin{eqnarray}
 \bar{g}_{4} &=&2W_{3}+(n-3)W_{4}-4W_{5}-2W_{6}-2(n-6)W_{7}+(n-5)W_{8} 
 \nonumber \\
 \bar{g}_{7} &=&\frac{1}{2}(n-3)W_{1}+\frac{1}{4}W_{2}+\frac{1}{4}
 (5n-16)W_{3}+\frac{1}{4}(n-3)(n-5)W_{4}  \nonumber \\
 &&+(n-3)(n-5)W_{5}+\frac{1}{4}(n-3)(n-6)W_{6}  \label{41} \\
 &&+\frac{1}{4}(n-3)(n-4)(n-11)W_{7}-\frac{1}{4}(n-3)(n-4)(n-5)W_{8}. 
 \nonumber \\
 \bar{g}_{8} &=&\frac{1}{4}W_{1}+\frac{3}{4}W_{3}+\frac{3}{8}(n-3)W_{4}+\frac{
 3}{2}(n-3)W_{5}+\frac{1}{2}(n-3)W_{6}  \nonumber \\
 &&+\frac{9}{8}(n-3)(n-4)W_{7}+\frac{1}{8}(n-3)(n-4)(n-5)W_{8}.  \nonumber
 \end{eqnarray}
 
 Hence, we have derived a Lagangean for the spin glass that is directly
 defined in terms of longitudinal, anomalous and replicon fields, and allows
 to perform a standard perturbation expansion. In summary, this field theory
 is defined by the quadratic mass term $\mathcal{L}^{(2)}$ in Eq. (31), the
 cubic interaction $\mathcal{L}^{(3)}$ in Eq. (38), and the bare propagators
 in Eq. (37), which involve the longitudinal, anomalous and replicon
 projectors given in Eqs. (18), (25) and (30), that verify the constraints in
 Eqs. (23) and (29). This in fact represents a general field theory that
 contains the spin glass symmetry and also other symmetries as will be
 discussed.

 \section{EQUATION OF STATE, BARE MASSES AND BARE COUPLINGS}

 The replica symmetric mean field value of the order parameter $Q$ is
 determined by the stationary condition $\mathcal{L}^{(1)}=0$, which from Eq.
 (5), provides the implicit form
 
 \begin{eqnarray}
 Q\Theta ^{-1} &=&\ll S^{\alpha }S^{\beta }\gg  \nonumber \\
 &=&\frac{{Tr}_{\left\{ S^{\alpha}\right\}}
\left[ S^{\alpha
 }S^{\beta }\exp \left( {\sum_{(\alpha \beta )} }QS^{\alpha
 }S^{\beta }+h{\sum_{\alpha } S^{\alpha }}\right) \right] }{
 {Tr}_{\left\{ S^{\alpha}\right\}}
\exp \left( 
 {\sum_{(\alpha \beta )} }QS^{\alpha }S^{\beta }+h
 {\sum_{\alpha } S^{\alpha }}\right) }  \label{42}
 \end{eqnarray}
 
 \noindent for the equation of state, $Q=Q(\Theta ,h,n)$.
 
 >From Eq. (8) we have the bare masses, (in units of $z$),
 
 \begin{eqnarray}
 M_{1} &=&-\Theta \left[ 1-\Theta ^{-1}-\ll S^{\alpha }S^{\beta }\gg
 ^{2}\right]  \nonumber \\
 M_{2} &=&-\Theta \left[ \ll S^{\alpha }S^{\beta }\gg -\ll S^{\alpha
 }S^{\beta }\gg ^{2}\right]  \label{43} \\
 M_{3} &=&-\Theta \left[ \ll S^{\alpha }S^{\beta }S^{\gamma }S^{\delta }\gg
 -\ll S^{\alpha }S^{\beta }\gg ^{2}\right]  \nonumber
 \end{eqnarray}
 
 \noindent and from Eq. (9) we have the bare cubic couplings, (in units of 
 $z^{3/2}$),
 
 \begin{eqnarray}
 W_{1} &=&\Theta ^{3/2}\left[ 1-3\ll S^{\alpha }S^{\beta }\gg ^{2}+2\ll
 S^{\alpha }S^{\beta }\gg ^{3}\right]  \nonumber \\
 W_{2} &=&\Theta ^{3/2}\left[ -2\ll S^{\alpha }S^{\beta }\gg +2\ll S^{\alpha
 }S^{\beta }\gg ^{3}\right]  \label{44} \\
 W_{3} &=&\Theta ^{3/2}\left[ -2\ll S^{\alpha }S^{\beta }\gg ^{2}+2\ll
 S^{\alpha }S^{\beta }\gg ^{3}\right]  \nonumber \\
 W_{4} &=&\Theta ^{3/2}\left[ 2\ll S^{\alpha }S^{\beta }\gg ^{3}-2\ll
 S^{\alpha }S^{\beta }\gg \ll S^{\alpha }S^{\beta }S^{\gamma }S^{\delta }\gg
 \right]  \nonumber \\
 W_{5} &=&\Theta ^{3/2}\left[ \ll S^{\alpha }S^{\beta }\gg \right. -2\ll
 S^{\alpha }S^{\beta }\gg ^{2}+2\ll S^{\alpha }S^{\beta }\gg ^{3}  \nonumber
 \\
 &&-\left. \ll S^{\alpha }S^{\beta }\gg \ll S^{\alpha }S^{\beta }S^{\gamma
 }S^{\delta }\gg \right]  \nonumber \\
 W_{6} &=&\Theta ^{3/2}\left[ -3\ll S^{\alpha }S^{\beta }\gg ^{2}+2\ll
 S^{\alpha }S^{\beta }\gg ^{3}+\ll S^{\alpha }S^{\beta }S^{\gamma }S^{\delta
 }\gg \right]  \nonumber \\
 W_{7} &=&\Theta ^{3/2}\left[ -\ll S^{\alpha }S^{\beta }\gg ^{2}\right. +2\ll
 S^{\alpha }S^{\beta }\gg ^{3}+\ll S^{\alpha }S^{\beta }S^{\gamma }S^{\delta
 }\gg  \nonumber \\
 &&-\left. 2\ll S^{\alpha }S^{\beta }\gg \ll S^{\alpha }S^{\beta }S^{\gamma
 }S^{\delta }\gg \right]  \nonumber \\
 W_{8} &=&\Theta ^{3/2}\left[ 2\ll S^{\alpha }S^{\beta }\gg ^{3}\right. -3\ll
 S^{\alpha }S^{\beta }\gg \ll S^{\alpha }S^{\beta }S^{\gamma }S^{\delta }\gg 
 \nonumber \\
 &&+\left. \ll S^{\alpha }S^{\beta }S^{\gamma }S^{\delta }S^{\mu }S^{\nu }\gg
 \right] .  \nonumber
 \end{eqnarray}
 
 \noindent The masses and the couplings, respectively in Eqs. (43) and (44),
 depend on the field $h$, and also the temperature $\Theta $, through $Q$.
 
 One finds that the spin correlations are given by 
 \begin{eqnarray}
 &\ll &S^{\alpha }S^{\beta }\gg =\overline{\tanh ^{2}(\sqrt{Q}y+h)}  \nonumber
 \\
 &\ll &S^{\alpha }S^{\beta }S^{\gamma }S^{\delta }\gg =\overline{\tanh ^{4}(
 \sqrt{Q}y+h)}  \label{45} \\
 &\ll &S^{\alpha }S^{\beta }S^{\gamma }S^{\delta }S^{\mu }S^{\nu }\gg =
 \overline{\tanh ^{6}(\sqrt{Q}y+h)}  \nonumber
 \end{eqnarray}
 
 \noindent with the notation
 
 \[
 \overline{\tanh ^{k}(\ldots )}=\frac{\int dy\;e^{-\frac{1}{2}y^{2}}\tanh
 ^{k}(\ldots )\cosh ^{n}(\ldots )}{\int dy\;e^{-\frac{1}{2}y^{2}}\cosh
 ^{n}(\ldots )}. 
 \]
 
 \noindent For small magnetic fields $h$, hence small $Q$, one gets, keeping
 terms up to order $Q^{3},h^{2}$ and $n$, 
 \begin{eqnarray}
 &\ll &S^{\alpha }S^{\beta }\gg =Q+(n-2)Q^{2}+\left( \frac{17}{3}-5n\right)
 Q^{3}+h^{2}+\ldots  \nonumber \\
 &\ll &S^{\alpha }S^{\beta }S^{\gamma }S^{\delta }\gg
 =3Q^{2}-2(10-3n)Q^{3}+\ldots  \label{46} \\
 &\ll &S^{\alpha }S^{\beta }S^{\gamma }S^{\delta }S^{\mu }S^{\nu }\gg
 =15Q^{3}+\ldots  \nonumber
 \end{eqnarray}
 
 Introducing the first of Eq. (46) into the stationary condition Eq. (42),
 gives
 
 \begin{equation}
 0=-tQ+(n-2)Q^{2}+\left( \frac{17}{3}-5n\right) Q^{3}+h^{2}+\ldots  \label{47}
 \end{equation}
 
 \noindent where we defined the temperature variable $t=\left( \Theta
 ^{-1}-1\right) $.
 
 Eqs. (32), (34), (43) and (46) give for the replicon, anomalous and
 longitudinal bare masses,

 \begin{eqnarray}
 m_{R} &=&(1+t)^{-1}\left[ t+2(1+t)Q-3Q^{2}+2(10-3n)Q^{3}+\ldots \right] 
 \nonumber \\
 m_{A} &=&(1+t)^{-1}\left[ t+(4-n)(1+t)Q-3(\allowbreak 3-n)Q^{2}\right. 
 \nonumber \\
 &&+\left. 2\left( 30-19n\right) Q^{3}+\ldots \right]  \label{48} \\
 m_{L} &=&(1+t)^{-1}\left[ t+2(2-n)(1+t)Q-\frac{1}{2}\left( 3\allowbreak
 (6-5n)\right. \right. +  \nonumber \\
 &&+\left. \left. n(1-n)(1+t)^{2}\right) Q^{2}+4\left( 15-17n\right)
 Q^{3}+\ldots \right]  \nonumber
 \end{eqnarray}
 
 \noindent and

 \begin{eqnarray}
 \bar{m}_{AL} &=&(1+t)^{-1}\left[ (1+t)Q+\frac{1}{2}\left( 3\left( n-3\right)
 +(1-n)(1+t)^{2}\right) Q^{2}\right.  \nonumber \\
 &&+\left. \left( 30-19n\right) Q^{3}+\ldots \right] .  \label{49}
 \end{eqnarray}
 
 \noindent Eqs. (39), (41), (44) and (46) give for the bare couplings, up to
 order $Q$,

 \begin{eqnarray}
 g_{1} &=&\left( 1+t\right) ^{-3/2}\left[ 1-3Q+\ldots \right] \allowbreak 
 \nonumber \\
 g_{2} &=&\left( 1+t\right) ^{-3/2}\left[ \allowbreak 4Q+\ldots \right]
 \label{50} \\
 g_{3} &=&\left( 1+t\right) ^{-3/2}\left[ -1+\allowbreak \left( 7-n\right)
 Q+\ldots \right]  \nonumber \\
 g_{4} &=&\left( 1+t\right) ^{-3/2}\left[ -1+\left( 7-2n\right) Q+\ldots
 \right]  \nonumber \\
 g_{5} &=&\left( 1+t\right) ^{-3/2}\left[ -\left( 1-\frac{1}{4}n\right)
 +\left( 7-\frac{11}{4}n\right) Q+\ldots \right]  \nonumber \\
 g_{6} &=&\left( 1+t\right) ^{-3/2}\left[ \left( 2-\frac{3}{4}n\right)
 -\left( 17-\frac{41}{4}n\right) Q+\ldots \right]  \nonumber \\
 g_{7} &=&\left( 1+t\right) ^{-3/2}(n-2)\left[ -\left( 1-\frac{1}{4}n\right) -
 \frac{1}{2}\left( 17-9n\right) Q+\ldots \right]  \nonumber \\
 g_{8} &=&\left( 1+t\right) ^{-3/2}n(n-1)\left[ -(2-n)+(17-15n)Q+\ldots
 \right]  \nonumber
 \end{eqnarray}
 
 \noindent and

 \begin{eqnarray}
 \bar{g}_{4} &=&\left( 1+t\right) ^{-3/2}\left[ -4Q+\ldots \right]  \nonumber
 \\
 \bar{g}_{7} &=&\left( 1+t\right) ^{-3/2}\left[ -\frac{1}{2}(3-n)+\frac{1}{2}
 \left( 29-8n\right) Q+\ldots \right]  \label{51} \\
 \bar{g}_{8} &=&\left( 1+t\right) ^{-3/2}\left[ \frac{1}{4}-\frac{1}{2}\left(
 9-3n\right) Q+\ldots \right] .  \nonumber
 \end{eqnarray}
 
 In zero magnetic field, $h=0$, Eq. (47) has the physical solutions, 
 $Q\geq 0$:

 \begin{eqnarray}
 Q &=&0\qquad \qquad \qquad \quad \mathrm{for}\quad t>0,  \label{52} \\
 Q &=&-\frac{1}{(2-n)}t+\ldots \quad \mathrm{for}\quad t<0.  \nonumber
 \end{eqnarray}
 
 \noindent One thus finds a spin glass transition in mean field at $t_{c}=0$.
 The three masses $m_{R}$, $m_{A}$, $m_{L}$ vanish at the transition point,
 the replicon, anomalous and longitudinal modes, becoming then simultaneously
 critical. Above the transition, $t>0$,
 
 \begin{equation}
 m_{R}=m_{A}=m_{L}\simeq t  \label{53}
 \end{equation}
 
 \noindent and, below the transition, $t<0$,
 
 \begin{eqnarray}
 m_{R} &\simeq &-\frac{n}{(2-n)}t  \nonumber \\
 m_{A} &\simeq &-\frac{2}{(2-n)}t  \label{54} \\
 m_{L} &\simeq &-t.  \nonumber
 \end{eqnarray}
 
 \noindent We have then that, above the transition the three masses are
 equal, vanishing at $t_{c}=0$, whereas below the transition an anisotropy
 develops, the replicon mass being zero while the anomalous and longitudinal
 masses are finite and degenerate in the limit $n\rightarrow 0$. For the
 couplings above the transition one has
 
 \begin{eqnarray}
 g_{1} &=&-g_{3}=-g_{4}=-\frac{4}{(4-n)}g_{5}=\frac{4}{(8-3n)}g_{6}  \nonumber
 \\
 &=&\frac{4}{(n-2)(n-4)}g_{7}=\frac{1}{n(n-1)(n-2)}g_{8}=1\quad  \label{55} \\
 g_{2} &=&0  \nonumber
 \end{eqnarray}
 
 \noindent and
 
 \begin{eqnarray}
 \bar{g}_{4} &=&0  \nonumber \\
 g_{1} &=&-\frac{2}{(3-n)}\bar{g}_{7}=4\bar{g}_{8}=1  \label{56}
 \end{eqnarray}
 
 In zero field and above the transition, where $Q=0$, there is indeed only
 one mass and one coupling, since
 
 \begin{eqnarray}
 M_{1} &=&t(1+t)^{-1},\quad M_{2}=M_{3}=0  \nonumber \\
 W_{1} &=&(1+t)^{-3/2},\quad W_{i}=0,\quad i=2,\ldots ,8.  \label{57}
 \end{eqnarray}
 
 In a finite magnetic field, $h\neq 0$, $Q$ is finite, which generates a
 splitting in the masses and the couplings. A spin glass transition occurs at
 the AT-line, $h=h_{c}(T)$, where the replicon modes become critical while
 the anomalous and longitudinal modes remain non-critical. The AT-line is
 then characterized by the vanishing of the replicon mass
 
 \begin{equation}
 m_{R}=0.  \label{58}
 \end{equation}
 
 \noindent The stationarity condition Eq. (47) together with Eqs. (48) and
 (58), determine the AT-line,
 
 \begin{equation}
 h^{2}=-\frac{1}{6}t^{3}-\frac{1}{4}nt^{2}.  \label{59}
 \end{equation}
 
 \noindent On the AT-line, $Q\simeq h^{2/3}$, and
 
 \begin{eqnarray}
 m_{A} &\simeq &\left( 1-\frac{n}{2}\right) h^{2/3}  \label{60} \\
 m_{L} &\simeq &\left( 1-n\right) h^{2/3}.  \nonumber
 \end{eqnarray}
 
 \noindent Likewise one can obtain the couplings from Eqs. (50)-(51).
 
 The expressions derived above for the bare masses and bare couplings provide
 the initial conditions for the renormalization group study.

 \section{RENORMALIZATION GROUP}

 We obtain the renormalization group equations by standard methods of
 integration of degrees of freedom over an infinitesimal momentum shell,
  $e^{-dl}\Lambda < \left|\mathbf{p}\right| < \Lambda$, at the cutoff $\Lambda$.$^{24}$
  This is the same renormalization scheme as used in Ref. 19. 
 The masses are determined by the zero momentum limit of the two-point vertex
 functions, $\Gamma _{i}^{(2)}(\mathbf{p})$, $i=R,A,L$.
 The leading, one-loop, order approximation for the
 two-point vertex functions corresponds to the ''bubble'' diagrams.
 The vertices are determined by the three-point vertex functions, 
 $\Gamma_{j}^{(3)}(\mathbf{p}_{1},\mathbf{p}_{2},\mathbf{p}_{3})$, 
 $j=1,\ldots,8$, calculated
 at zero momenta, with the definition:
   $\Gamma _{1}^{(3)}=\Gamma _{RRR,1}$, $\Gamma
 _{2}^{(3)}=\Gamma _{RRR,2}$, $\Gamma _{3}^{(3)}=\Gamma _{RRA}$, $\Gamma
 _{4}^{(3)}=\Gamma _{RRL}$, $\Gamma _{5}^{(3)}=\Gamma _{RAA}$, $\Gamma
 _{6}^{(3)}=\Gamma _{AAA}$, $\Gamma _{7}^{(3)}=\Gamma _{AAL}$, $\Gamma
 _{8}^{(3)}=\Gamma _{LLL}$. The leading, one-loop, order
 approximation for the three-point vertex functions corresponds to the
 ''triangle'' diagrams.
 
 The renormalization equations are given, to lowest order in
 $\varepsilon =6-d$, by
 
 \begin{eqnarray}
 \frac{\partial m_{R}}{\partial l} &=&(2-\eta _{R})m_{R}-\left[ g_{1}^{2}
 \frac{(n^{4}-8n^{3}+19n^{2}-4n-16)}{(n-1)(n-2)^{2}}\right.  \label{67} \\
 &&+g_{1}g_{2}\frac{2(3n^{2}-15n+16)}{(n-1)(n-2)^{2}}+\left. g_{2}^{2}\frac{
 (n^{3}-9n^{2}+26n-22)}{2(n-1)(n-2)^{2}}\right] I_{RR}  \nonumber \\
 &&-g_{3}^{2}\left[ \frac{8(n-1)(n-4)}{n(n-2)^{2}}\right]
 I_{RA}-g_{4}^{2}\left[ \frac{8}{n(n-1)}\right] I_{RL}  \nonumber \\
 &&-g_{5}^{2}\left[ \frac{16}{(n-2)^{2}}\right] I_{AA}  \nonumber
 \end{eqnarray}
 
 \begin{eqnarray}
 \frac{\partial m_{A}}{\partial l} &=&(2-\eta _{A})m_{A}-g_{3}^{2}\left[ 
 \frac{2(n-3)(n-4)}{(n-2)^{2}}\right] I_{RR}  \label{68} \\
 &&-g_{5}^{2}\left[ \frac{16n(n-3)}{(n-1)(n-2)^{2}}\right]
 I_{RA}-g_{6}^{2}\left[ \frac{32}{n(n-2)^{2}}\right] I_{AA}  \nonumber \\
 &&-g_{7}^{2}\left[ \frac{32}{n(n-1)(n-2)^{2}}\right] I_{AL}  \nonumber
 \end{eqnarray}
 
 \begin{eqnarray}
 \frac{\partial m_{L}}{\partial l} &=&(2-\eta _{L})m_{L}-g_{4}^{2}\left[ 
 \frac{2(n-3)}{(n-1)}\right] I_{RR}  \label{69} \\
 &&-g_{7}^{2}\left[ \frac{16}{n(n-2)^{2}}\right] I_{AA}-g_{8}^{2}\left[ \frac{
 4}{n^{3}(n-1)^{3}}\right] I_{LL}  \nonumber
 \end{eqnarray}
 
 \begin{eqnarray}
 \frac{\partial g_{1}}{\partial l} &=&\frac{1}{2}(\varepsilon -3\eta
 _{R})g_{1}+\left[ g_{1}^{3}\frac{(n^{5}-10n^{4}+33n^{3}-8n^{2}-104n+112)}{
 (n-1)(n-2)^{3}}\right.  \label{70} \\
 &&+g_{1}^{2}g_{2}\frac{3(3n^{3}-27n^{2}+64n-48)}{(n-1)(n-2)^{3}}
 +g_{1}g_{2}^{2}\frac{3(-n^{3}+8n^{2}-17n+12)}{(n-1)(n-2)^{3}}  \nonumber \\
 &&\left. -g_{2}^{3}\frac{1}{(n-2)^{3}}\right] I_{RRR}+\left[ g_{2}g_{3}^{2}
 \frac{12}{(n-2)^{3}}\right.  \nonumber \\
 &&+g_{1}g_{3}^{2}\left. \frac{12(n^{3}-9n^{2}+20n-16)}{n(n-2)^{3}}\right]
 I_{RRA}+g_{1}g_{4}^{2}\left[ \frac{24}{n(n-1)}\right] I_{RRL}  \nonumber \\
 &&-g_{3}^{2}g_{5}\left[ \frac{48}{(n-2)^{3}}\right] I_{RAA}+g_{5}^{3}\left[ 
 \frac{64}{(n-2)^{3}}\right] I_{AAA}  \nonumber
 \end{eqnarray}
 
 \begin{eqnarray}
 \frac{\partial g_{2}}{\partial l} &=&\frac{1}{2}(\varepsilon -3\eta
 _{R})g_{2}-\left[ g_{1}^{3}\frac{24n}{(n-2)^{2}}\right.  \label{71} \\
 &&+g_{1}^{2}g_{2}\frac{6(n^{3}-5n^{2}-8n+16)}{(n-1)(n-2)^{2}}-g_{1}g_{2}^{2}
 \frac{3(6n^{2}-38n+40)}{(n-1)(n-2)^{2}}  \nonumber \\
 &&-g_{2}^{3}\left. \frac{(n^{3}-11n^{2}+38n-34)}{(n-1)(n-2)^{2}}\right]
 I_{RRR}+\left[ g_{1}g_{3}^{2}\frac{96}{(n-2)^{2}}\right.  \nonumber \\
 &&+\left. g_{2}g_{3}^{2}\frac{24(n^{2}-6n+4)}{n(n-2)^{2}}\right]
 I_{RRA}+g_{2}g_{4}^{2}\left[ \frac{24}{n(n-1)}\right] I_{RRL}  \nonumber \\
 &&+g_{3}^{2}g_{5}\left[ \frac{96}{(n-2)^{2}}\right] I_{RAA}  \nonumber
 \end{eqnarray}
 
 \begin{eqnarray}
 \frac{\partial g_{3}}{\partial l} &=&\frac{1}{2}(\varepsilon -2\eta
 _{R}-\eta _{A})g_{3}  \label{72} \\
 &&+\left[ g_{1}^{2}g_{3}\frac{(n^{5}-12n^{4}+47n^{3}-44n^{2}-48n+64)}{
 (n-1)(n-2)^{3}}\right.  \nonumber \\
 &&+g_{2}^{2}g_{3}\frac{(n^{4}-12n^{3}+50n^{2}-81n+44)}{(n-1)(n-2)^{3}} 
 \nonumber \\
 &&+\left. g_{1}g_{2}g_{3}\frac{(14n^{3}-102n^{2}+208n-128)}{(n-1)(n-2)^{3}}
 \right] I_{RRR}  \nonumber \\
 &&-\left[ g_{1}g_{3}g_{5}\frac{8n(n^{2}-9n+12)}{(n-1)(n-2)^{3}}\right.
 -g_{2}g_{3}g_{5}\frac{8n(n^{2}-6n+7)}{(n-1)(n-2)^{3}}  \nonumber \\
 &&-\left. g_{3}^{3}\frac{4(2n3-15n^{2}+28n-16)}{n(n-2)^{3}}\right]
 I_{RRA}+g_{3}g_{4}^{2}\left[ \frac{8}{n(n-.1)}\right] I_{RRL}  \nonumber \\
 &&+\left[ g_{3}g_{5}^{2}\frac{32n}{(n-2)^{3}}+g_{3}^{2}g_{6}\frac{
 16(n^{2}-7n+8)}{n(n-2)^{3}}\right] I_{RAA}  \nonumber \\
 &&+g_{3}g_{4}g_{7}\left[ \frac{32}{n(n-1)(n-2)}\right]
 I_{RAL}+g_{5}^{2}g_{6}\left[ \frac{64}{(n-2)^{3}}\right] I_{AAA}  \nonumber
 \end{eqnarray}
 
 \begin{eqnarray}
 \frac{\partial g_{4}}{\partial l} &=&\frac{1}{2}(\varepsilon -2\eta
 _{R}-\eta _{L})g_{4}+\left[ g_{1}^{2}g_{4}\frac{
 (2n^{4}-16n^{3}+38n^{2}-8n-32)}{(n-1)(n-2)^{2}}\right.  \label{73} \\
 &&+g_{2}^{2}g_{4}\frac{(n^{3}-9n^{2}+26n-22)}{(n-1)(n-2)^{2}}
 +g_{1}g_{2}g_{4}\left. \frac{(12n^{2}-60n+64)}{(n-1)(n-2)^{2}}\right] I_{RRR}
 \nonumber \\
 &&+g_{3}^{2}g_{4}\left[ \frac{8(n^{2}-5n+4)}{n(n-2)^{2}}\right]
 I_{RRA}+g_{4}^{3}\left[ \frac{8}{n(n-1)}\right] I_{RRL}  \nonumber \\
 &&+g_{3}^{2}g_{7}\left[ \frac{16(n^{2}-5n+4)}{n(n-2)^{3}}\right]
 I_{RAA}+g_{4}^{2}g_{8}\left[ \frac{8}{n^{2}(n-1)^{2}}\right] I_{RLL} 
 \nonumber \\
 &&+g_{5}^{2}g_{7}\left[ \frac{64}{(n-2)^{3}}\right] I_{AAA}  \nonumber
 \end{eqnarray}
 
 \begin{eqnarray}
 \frac{\partial g_{5}}{\partial l} &=&\frac{1}{2}(\varepsilon -\eta
 _{R}-2\eta _{A})g_{5}-\left[ g_{1}g_{3}^{2}\frac{(n^{3}-13n^{2}+48n-48)}{
 (n-2)^{3}}\right.  \label{74} \\
 &&-\left. g_{2}g_{3}^{2}\frac{n^{3}-10n^{2}+31n-28}{(n-2)^{3}}\right] I_{RRR}
 \nonumber \\
 &&+\left[ g_{1}g_{5}^{2}\frac{4(n^{4}-8n^{3}+19n^{2}-4n-16)}{(n-1)(n-2)^{3}}
 \right. +g_{2}g_{5}^{2}\frac{2(6n^{2}-30n+32)}{(n-1)(n-2)^{3}}  \nonumber \\
 &&+\left. g_{3}^{2}g_{5}\frac{8(n^{2}-5n+4)}{(n-2)^{3}}\right]
 I_{RRA}+\left[ g_{5}^{3}\frac{16(n^{2}-3n+4)}{(n-1)(n-2)^{3}}\right. 
 \nonumber \\
 &&+\left. g_{3}g_{5}g_{6}\frac{32(n^{2}-5n+4)}{n(n-2)^{3}}\right]
 I_{RAA}+g_{4}g_{5}g_{7}\left[ \frac{32}{n(n-1)(n-2)}\right] I_{RAL} 
 \nonumber \\
 &&-g_{5}g_{6}^{2}\left[ \frac{64}{n(n-2)^{3}}\right]
 I_{AAA}+g_{5}g_{7}^{2}\left[ \frac{32}{n(n-1)(n-2)^{2}}\right] I_{AAL} 
 \nonumber
 \end{eqnarray}
 
 \begin{eqnarray}
 \frac{\partial g_{6}}{\partial l} &=&\frac{1}{2}(\varepsilon -3\eta
 _{A})g_{6}+g_{3}^{3}\left[ \frac{(n^{4}-14n^{3}+69n^{2}-140n+96)}{(n-2)^{3}}
 \right] I_{RRR}  \label{75} \\
 &&+g_{3}g_{5}^{2}\left[ \frac{12n(n-3)(n-4)}{(n-2)^{3}}\right]
 I_{RRA}-g_{5}^{2}g_{6}\left[ \frac{48n(n-3)}{(n-1)(n-2)^{3}}\right] I_{RAA} 
 \nonumber \\
 &&+g_{6}^{3}\left[ \frac{64(n-3)}{n(n-2)^{3}}\right]
 I_{AAA}+g_{6}g_{7}^{2}\left[ \frac{96}{n(n-1)(n-2)^{2}}\right] I_{AAL} 
 \nonumber
 \end{eqnarray}
 
 \begin{eqnarray}
 \frac{\partial g_{7}}{\partial l} &=&\frac{1}{2}(\varepsilon -2\eta
 _{A}-\eta _{L})g_{7}+g_{3}^{2}g_{4}\left[ \frac{2(n^{2}-7n+12)}{(n-2)}
 \right] I_{RRR}  \label{76} \\
 &&+g_{4}g_{5}^{2}\left[ \frac{8n(n-3)}{(n-1)(n-2)}\right]
 I_{RRA}+g_{5}^{2}g_{7}\left[ \frac{16n(n-3)}{(n-1)(n-2)^{2}}\right] I_{RAA} 
 \nonumber \\
 &&+g_{6}^{2}g_{7}\left[ \frac{64}{n(n-2)^{2}}\right] I_{AAA}+g_{7}^{3}\left[ 
 \frac{32}{n(n-1)(n-2)^{2}}\right] I_{AAL}  \nonumber \\
 &&+g_{7}^{2}g_{8}\left[ \frac{16}{n^{2}(n-1)^{2}(n-2)}\right] I_{ALL} 
 \nonumber
 \end{eqnarray}
 
 \begin{eqnarray}
 \frac{\partial g_{8}}{\partial l} &=&\frac{1}{2}(\varepsilon -3\eta
 _{L})g_{8}+g_{4}^{3}\left[ 4n(n-3)\right] I_{RRR}  \label{77} \\
 &&+g_{7}^{3}\left[ \frac{64(n-1)}{(n-2)^{3}}\right] I_{AAA}+g_{8}^{3}\left[ 
 \frac{8}{n^{3}(n-1)^{3}}\right] I_{LLL}  \nonumber
 \end{eqnarray}
 with
 
 \begin{eqnarray}
 \eta _{R} &=&\frac{1}{3}\left\{ \left[ g_{1}^{2}\frac{
 (n^{4}-8n^{3}+19n^{2}-4n-16)}{(n-1)(n-2)^{2}}\right. \right.  \label{78} \\
 &&+g_{1}g_{2}\frac{2(3n^{2}-15n+16)}{(n-1)(n-2)^{2}}+g_{2}^{2}\left. \frac{
 (n^{3}-9n^{2}+26n-22)}{2(n-1)(n-2)^{2}}\right] T_{RR}  \nonumber \\
 &&+g_{3}^{2}\left[ \frac{4(n-1)(n-4)}{n(n-2)^{2}}\right] (T_{RA}+T_{AR}) 
 \nonumber \\
 &&+g_{4}^{2}\left[ \frac{4}{n(n-1)}\right] (T_{RL}+T_{LR})+\left.
 g_{5}^{2}\left[ \frac{16}{(n-2)^{2}}\right] T_{AA}\right\}  \nonumber
 \end{eqnarray}
 
 \begin{eqnarray}
 \eta _{A} &=&\frac{1}{3}\left\{ g_{3}^{2}\left[ \frac{2(n-3)(n-4)}{(n-2)^{2}}
 \right] T_{RR}\right.  \label{79} \\
 &&+g_{5}^{2}\left[ \frac{8n(n-3)}{(n-1)(n-2)^{2}}\right]
 (T_{RA}+T_{AR})+g_{6}^{2}\left[ \frac{32}{n(n-2)^{2}}\right] T_{AA} 
 \nonumber \\
 &&+\left. g_{7}^{2}\left[ \frac{16}{n(n-1)(n-2)^{2}}\right]
 (T_{AL}+T_{LA})\right\}  \nonumber
 \end{eqnarray}
 
 \begin{eqnarray}
 \eta _{L} &=&\frac{1}{3}\left\{ g_{4}^{2}\left[ \frac{2(n-3)}{(n-1)}\right]
 T_{RR}\right. +g_{7}^{2}\left[ \frac{16}{n(n-2)^{2}}\right] T_{AA}
 \label{80} \\
 &&+\left. g_{8}^{2}\left[ \frac{4}{n^{3}(n-1)^{3}}\right] T_{LL}\right\} 
 \nonumber
 \end{eqnarray}
 
 \noindent defining
 
 \[
 I_{ij}=\frac{1}{(1+m_{i})(1+m_{j})}, 
 \]
 
 \[
 I_{ijk}=\frac{1}{(1+m_{i})(1+m_{j})(1+m_{k})}, 
 \]
 
 \[
 T_{ij}=\frac{1+3m_{j}}{(1+m_{i})(1+m_{j})^{3}}. 
 \]
 
 \noindent In Eqs. (61)-(74) 
 the usual geometrical factor\textbf{\ }$K_{d}=2/(4\pi
 )^{d/2}\Gamma (d/2)$\textbf{\ }is absorbed in the couplings, i.e.,\textbf{\ }
 $K_{d}^{1/2}g_{i}\rightarrow g_{i}$. The one-loop
 perturbation equations for the two-point and three-point vertex functions
 contain $1/n$ - factors, which consequently appear in the renormalization
 equations. Those factors, which come always associated with terms involving
 the anomalous and longitudinal modes, could in principle, lead to
 divergences in the limit $n\rightarrow 0$. The spin glass symmetry does
 however prevent this problem, as will be seen.
 
 As mentioned before, the field theory considered, and hence the
 renormalization equations just derived, are quite general, containing
 different symmetries, which depend on $n$. The spin glass is obtained in the
 limit $n\rightarrow 0$. Here we assume that perturbation theory works in a
 regular way as presented above. The possibility that subdominant dangerous
 terms may take over (with as a consequence a breakdown of the symmetry
 involved here) will be considered separately\textbf{.}$^{20}$ With that
 assumption there are now two important features. First, at the mean-field
 level, i.e. in 0-loop order, the spin glass is characterized by a degeneracy
 in the masses and the couplings, i.e. $m_{A}$ and $m_{L}$, $g_{3}$ and $
 g_{4} $, $g_{6}$ and $g_{7}$, $g_{7}$ and $g_{8}/n$, are equal, when $
 n\rightarrow 0$; we then defined the variables, $\bar{m}_{AL}$, $\bar{g}_{4}$
 , $\bar{g}_{7}$ and $\bar{g}_{8}$, in Eqs. (33) and (40), which although
 incorporating $1/n$ - factors, are in 0-loop order, well defined quantities
 in terms of the spin correlations functions, when $n\rightarrow 0$, see Eqs
 (34) and (41). Secondly, starting with the set of variables $m_{R}$, $m_{A}$
 , $\bar{m}_{AL}$, $g_{1}$, $g_{2}$, $g_{3}$, $\bar{g}_{4}$, $g_{5}$, $g_{6}$
 , $\bar{g}_{7}$, $\bar{g}_{8}$, one can built a set of equations for the
 1-loop two-point and three-point vertex functions corresponding to those
 variables (defining $\bar{\Gamma}_{AL}=(\Gamma _{A}-\Gamma _{L})/n$, $\bar{
 \Gamma}_{4}=4(\Gamma _{4}-\Gamma _{3})/n$, $\bar{\Gamma}_{7}=2(\Gamma
 _{7}-\Gamma _{6})/n$, $\bar{\Gamma}_{8}=(\Gamma _{8}-3n\Gamma _{7}+2n\Gamma
 _{6})/n^{3}$), which is free of $1/n$ - factors, and hence is well behaved
 in the limit $n\rightarrow 0$. This means that the degeneracy that exists,
 in the masses and the couplings in 0-loop order, is also present in 1-loop
 order, and consequently, the renormalization group transformation preserves
 this symmetry. One can then write the renormalization equations for the new
 set of variables, and $n=0$:
 
 \begin{eqnarray}
 \frac{\partial m_{R}}{\partial l} &=&(2-\eta _{R})m_{R}-\left[
 4g_{1}^{2}-8g_{1}g_{2}+\frac{11}{4}g_{2}^{2}\right] I_{RR}  \label{81} \\
 &&+\left[ 10g_{3}^{2}+4g_{3}\bar{g}_{4}\right]
 I_{RA}-4g_{5}^{2}I_{AA}+8g_{3}^{2}\bar{m}_{AL}I_{RAA}  \nonumber
 \end{eqnarray}
 
 \begin{equation}
 \frac{\partial m_{A}}{\partial l}=(2-\eta
 _{A})m_{A}-6g_{3}^{2}I_{RR}+8\left[ g_{6}^{2}+g_{6}\bar{g}_{7}\right]
 I_{AA}+8g_{6}^{2}\bar{m}_{AL}I_{AAA}  \label{82}
 \end{equation}
 
 \begin{eqnarray}
 \frac{\partial \bar{m}_{AL}}{\partial l} &=&(2-\eta _{A})\bar{m}_{AL}-\bar{
 \eta}_{AL}m_{A}+\frac{3}{2}\left[ g_{3}^{2}+2g_{3}\bar{g}_{4}\right] I_{RR}
 \label{83} \\
 &&-12g_{5}^{2}I_{RA}-\left[ 5g_{6}^{2}\right. +16g_{6}\bar{g}_{7}+8g_{6}\bar{
 g}_{8}+\left. 6\bar{g}_{7}^{2}\right] I_{AA}  \nonumber \\
 &&-8\left[ g_{6}^{2}+\allowbreak 2g_{6}\bar{g}_{7}\right] \bar{m}_{AL}I_{AAA}
 \nonumber
 \end{eqnarray}
 
 \begin{eqnarray}
 \frac{\partial g_{1}}{\partial l} &=&\frac{1}{2}(\varepsilon -3\eta
 _{R})g_{1}+\left[ 14g_{1}^{3}-18g_{1}^{2}g_{2}+\frac{9}{2}g_{1}g_{2}^{2}+
 \frac{1}{8}g_{2}^{3}\right] I_{RRR}  \label{84} \\
 &&-\left[ 18g_{1}g_{3}^{2}+\frac{3}{2}g_{2}g_{3}^{2}+12g_{1}g_{3}\bar{g}
 _{4}\right] I_{RRA}+6g_{3}^{2}g_{5}I_{RAA}-8g_{5}^{3}I_{AAA}  \nonumber
 \end{eqnarray}
 
 \begin{eqnarray}
 \frac{\partial g_{2}}{\partial l} &=&\frac{1}{2}(\varepsilon -3\eta
 _{R})g_{2}+\left[ 24g_{1}^{2}g_{2}-30g_{1}g_{2}^{2}+\frac{17}{2}
 g_{2}^{3}\right] I_{RRR}  \label{85} \\
 &&+\left[ 24g_{1}g_{3}^{2}-36g_{2}g_{3}^{2}-12g_{2}g_{3}\bar{g}_{4}\right]
 I_{RRA}+24g_{3}^{2}g_{5}I_{RAA}  \nonumber
 \end{eqnarray}
 
 \begin{eqnarray}
 \frac{\partial g_{3}}{\partial l} &=&\frac{1}{2}(\varepsilon -2\eta
 _{R}-\eta _{A})g_{3}+\left[ 8g_{1}^{2}g_{3}+\frac{11}{2}
 g_{2}^{2}g_{3}-16g_{1}g_{2}g_{3}\right] I_{RRR}  \label{86} \\
 &&-\left[ 10g_{3}^{3}+4g_{3}^{2}\bar{g}_{4}\right] I_{RRA}+\left[
 14g_{3}^{2}g_{6}+8g_{3}^{2}\bar{g}_{7}+4g_{3}\bar{g}_{4}g_{6}\right] I_{RAA}
 \nonumber \\
 &&-8g_{5}^{2}g_{6}I_{AAA}  \nonumber
 \end{eqnarray}
 
 \begin{eqnarray}
 \frac{\partial \bar{g}_{4}}{\partial l} &=&\frac{1}{2}(\varepsilon -2\eta
 _{R}-\eta _{A})\bar{g}_{4}+2\bar{\eta}_{AL}g_{3}+\left[
 16g_{1}^{2}g_{3}+8g_{1}^{2}\bar{g}_{4}+\frac{7}{2}g_{2}^{2}g_{3}\right.
 \label{87} \\
 &&-12g_{1}g_{2}g_{3}+\left. \frac{11}{2}g_{2}^{2}\bar{g}_{4}-16g_{1}g_{2}
 \bar{g}_{4}\right] I_{RRR}+\left[ 48g_{1}g_{3}g_{5}\right. -28g_{2}g_{3}g_{5}
 \nonumber \\
 &&-2g_{3}^{3}-10g_{3}^{2}\bar{g}_{4}-\left. 4g_{3}\bar{g}_{4}\right]
 I_{RRA}+\left[ 16g_{3}g_{5}^{2}\right. +8g_{3}g_{6}^{2}+44g_{3}^{2}\bar{g}
 _{7}  \nonumber \\
 &&+8g_{3}\bar{g}_{4}g_{6}+32g_{3}^{2}\bar{g}_{8}+16g_{3}\bar{g}
 _{4}\bar{g}_{7}+\left. 2\bar{g}_{4}^{2}g_{6}\right] I_{RAA}-16g_{5}^{2}\bar{g}
 _{7}I_{AAA}  \nonumber
 \end{eqnarray}
 
 \begin{eqnarray}
 \frac{\partial g_{5}}{\partial l} &=&\frac{1}{2}(\varepsilon -\eta
 _{R}-2\eta _{A})g_{5}-\left[ 6g_{1}g_{3}^{2}-\frac{7}{2}g_{2}g_{3}^{2}
 \right] I_{RRR}+\left[ -8g_{1}g_{5}^{2}\right.  \label{88} \\
 &&+8g_{2}g_{5}^{2}-4\left. g_{3}^{2}g_{5}\right] I_{RRA}+\left[
 8g_{5}^{3}\right. +20g_{3}g_{5}g_{6}+8g_{3}g_{5}\bar{g}_{7}  \nonumber \\
 &&+4\left. \bar{g}_{4}g_{5}g_{6}\right] I_{RAA}-\left[
 4g_{5}g_{6}^{2}+8g_{5}g_{6}\bar{g}_{7}\right] I_{AAA}  \nonumber
 \end{eqnarray}
 
 \begin{equation}
 \frac{\partial g_{6}}{\partial l}=\frac{1}{2}(\varepsilon -3\eta
 _{A})g_{6}-12g_{3}^{3}I_{RRR}-\left[ 20g_{6}^{3}+24g_{6}^{2}\bar{g}
 _{7}\right] I_{AAA}  \label{89}
 \end{equation}
 
 \begin{eqnarray}
 \frac{\partial \bar{g}_{7}}{\partial l} &=&\frac{1}{2}(\varepsilon -3\eta
 _{A})\bar{g}_{7}+\bar{\eta}_{AL}g_{6}+\left[ 3g_{3}^{3}-6g_{3}^{2}\bar{g}
 _{4}\right] I_{RRR}+12g_{3}g_{5}^{2}I_{RRA}  \label{90} \\
 &&-12g_{5}^{2}g_{6}I_{RAA}-\left[ 4g_{6}^{3}+36g_{6}^{2}\bar{g}_{7}+28g_{6}
 \bar{g}_{7}^{2}+16g_{6}^{2}\bar{g}_{8}\right] I_{AAA}  \nonumber
 \end{eqnarray}
 
 \begin{eqnarray}
 \frac{\partial \bar{g}_{8}}{\partial l} &=&\frac{1}{2}(\varepsilon -3\eta
 _{A})\bar{g}_{8}+\frac{3}{2}\bar{\eta}_{AL}\bar{g}_{7}+\frac{3}{4}\left[
 g_{3}^{3}+3g_{3}^{2}\bar{g}_{4}-3g_{3}\bar{g}_{4}^{2}\right] I_{RRR}
 \label{91} \\
 &&+9\left[ g_{3}g_{5}^{2}\right. +\left. \bar{g}_{4}g_{5}^{2}\right]
 I_{RRA}+18\left[ g_{5}^{2}g_{6}-g_{5}^{2}\bar{g}_{7}\right] I_{RAA}-\left[
 12g_{6}^{2}\bar{g}_{7}\right.  \nonumber \\
 &&+g_{6}^{3}+42g_{6}\bar{g}_{7}^{2}+14\bar{g}_{7}^{3}+12g_{6}^{2}\bar{g}
 _{8}+48\left. g_{6}\bar{g}_{7}\bar{g}_{8}\right] I_{AAA}  \nonumber
 \end{eqnarray}
 
 \noindent with
 
 \begin{eqnarray}
 \eta _{R} &=&\frac{1}{3}\left\{ \left[ 4g_{1}^{2}-8g_{1}g_{2}+\frac{11}{4}
 g_{2}^{2}\right] T_{RR}\right.  \label{92} \\
 &&-\left[ 5g_{3}^{2}+2g_{3}\bar{g}_{4}\right] (T_{RA}+T_{AR})+\left.
 4g_{5}^{2}T_{AA}\right\}  \nonumber
 \end{eqnarray}
 
 \begin{equation}
 \eta _{A}=\frac{1}{3}\left\{ 6g_{3}^{2}T_{RR}-8\left[ g_{6}^{2}+g_{6}\bar{g}
 _{7}\right] T_{AA}\right\}  \label{93}
 \end{equation}
 
 \begin{eqnarray}
 \bar{\eta}_{AL} &=&-\frac{1}{2}\left[ g_{3}^{2}+2g_{3}\bar{g}_{4}\right]
 T_{RR}+2g_{5}^{2}(T_{RA}+T_{AR})  \label{94} \\
 &&+\frac{1}{3}\left[ 5g_{6}^{2}+16g_{6}\bar{g}_{7}+8g_{6}\bar{g}_{8}+6\bar{g}
 _{7}^{2}\right] T_{AA}.  \nonumber
 \end{eqnarray}
 
 We remark that the theory with $n$ finite, allows the three masses, as well
 as the eight couplings $g_{j}$, to be different, while the theory with $n=0$
 , imposes $m_{A}=m_{L}$ and $g_{3}=g_{4}$, $g_{6}=g_{7}=g_{8}/n$. The two
 sets of renormalization equations, Eqs (61)-(74) and Eqs. (75)-(88), imply
 two different procedures, i.e. the first corresponds to keeping $n$ finite
 along the renormalization iteration, and only at the end set it zero, while
 the second corresponds to imposing $n=0$ all the way along the
 renormalization iteration. The two procedures may lead to different results,
 as $n$ influences the flow equations. When using the first procedure to
 study the spin glass one has to carefully take into account the symmetry of
 the problem, in order to obtain the relevant results. The latter procedure
 naturally enforces the spin glass symmetry. We will discuss the cases of
 zero and small magnetic field using the two procedures, for illustration.
 Let us then look for the fixed-points (f.p.) of the renormalization group
 equations.
 
 In zero magnetic field, $h=0$, and $t>0$, the three masses, $m_{i}$, 
 $i=R,A,L $, are equal, see Eq. (53), and the eight couplings, $g_{j},$ 
 $j=1,\ldots ,8$ are simply related, as in Eqs. (55)-(56). So, if one imposes
 this symmetry 
 \begin{equation}
 m_{R}=m_{A}=m_{L}=m  \label{95}
 \end{equation}
 \begin{eqnarray}
 g_{1} &=&-g_{3}=-g_{4}=\frac{4}{(n-4)}g_{5}=\frac{4}{(8-3n)}g_{6}  \nonumber
 \\
 &=&\frac{4}{(n-2)(n-4)}g_{7}=\frac{1}{n(n-1)(n-2)}g_{8}=g  \label{96} \\
 g_{2} &=&0.  \nonumber
 \end{eqnarray}
 
 \noindent on the renormalization equations, Eqs. (61)-(74), they reduce to
 
 \begin{equation}
 \frac{\partial m}{\partial l}=(2-\eta )m+\left( 2-n\right) g^{2}\frac{1}{
 (1+m)^{2}}  \label{97}
 \end{equation}
 
 \begin{equation}
 \frac{\partial g}{\partial l}=\frac{1}{2}(\varepsilon -3\eta )g-\left(
 2-n\right) g^{3}\frac{1}{(1+m)^{3}}  \label{98}
 \end{equation}
 
 \noindent with 
 \begin{equation}
 \eta =-\frac{1}{3}(2-n)g^{2}\allowbreak \frac{1+3m}{(1+m)^{4}}.  \label{99}
 \end{equation}
 
 \noindent Equivalently, if one imposes the zero-field symmetry, with $n=0$,
 
 \begin{equation}
 m_{R}=m_{A}=m,\quad \bar{m}_{AL}=0
 \end{equation}
 
 \begin{eqnarray}
 g_{1} &=&-g_{3}=-g_{5}=\frac{1}{2}g_{6}=-\frac{2}{3}\bar{g}_{7}=4\bar{g}
 _{8}=g  \label{101} \\
 g_{2} &=&\bar{g}_{4}=0  \nonumber
 \end{eqnarray}
 on the renormalization equations, Eqs.(75)-(88), they reduce to Eqs.
 (91)-(93), naturally with $n=0$. For the Eqs. (91)-(93), one finds:
 
 $(i)$ the trivial Gaussian f.p.
 
 \begin{equation}
 m^{*}=0,\mathrm{\quad }g^{*}=0,  \label{102}
 \end{equation}
 
 \noindent which is unstable for $\varepsilon >0$, with eigenvalues $\lambda
 _{1}^{G}=2$, $\lambda _{2}^{G}=\frac{1}{2}\varepsilon $;
 
 $(ii)$ the nontrivial zero-field f.p. 
 \begin{equation}
 m^{*}=-\frac{\varepsilon }{2},\mathrm{\quad }g^{*2}=\frac{\varepsilon }{\left(
 2-n\right) },  \label{103}
 \end{equation}
 
 \noindent which is stable for $\varepsilon >0$, with eigenvalues $\lambda
 _{1}^{ZF}=2(1-\varepsilon )$, $\lambda _{2}^{ZF}=-\varepsilon $. The
 zero-field f.p. has then the associated critical exponents 
 \begin{equation}
 \nu =\frac{1}{2}+\frac{5}{12}\varepsilon ,\quad \eta =-\frac{1}{3}\varepsilon
 \label{104}
 \end{equation}
 
 \noindent in agreement with the results of Harris \textit{et al}.$^{18}$
 
 In the generalized parameter space, with the renormalization equations
 (61)-(74), or (75)-(88), one can get the fixed-points associated to
 zero-field by searching for fixed-points with the same symmetry as in Eqs.
 (89)-(90), or (94)-(95), and indeed one then finds two fixed-points, which
 correspond to Eqs. (96)-(97), and $\bar{m}_{AL}^{*}=0$. For $\varepsilon
 >0 $, the Gaussian f.p. is unstable in all the directions, while the
 non-trivial zero-field f.p. is stable in two but unstable in six directions
 in the coupling-space, see Table I, concerning Eqs. (75)-(88). The
 appearance of unstable directions for the couplings in this situation,
 arises from the fact that the generalized parameter space has a lower
 symmetry than the characteristic of zero-field.$^{24}$ However, of crucial
 importance is the fact that the initial values of the couplings, Eqs.
 (55)-(56), lie along one of the stable directions, i.e. Eqs. (61)-(74) and
 Eqs. (75)-(88), have an eigenvector (associated to the eigenvalue $-1$),
 which has precisely the symmetry in, respectively, Eq. (90) and Eq. (95),
 see Table I for the latter. So, the system will start and will remain in a
 stable direction, the unstable directions playing no role. That stable
 direction in fact represents an invariant manifold.
 
 In a nonzero magnetic field, $h\neq 0$, an anisotropy develops in the
 masses, and in the couplings. To discuss the effect of a small magnetic
 field, $h\ll 1$, we linearize the renormalization equations about the
 zero-field f.p. For the mass equations, Eqs. (61)-(63), we obtain the
 eigenvalues
 
 \begin{eqnarray}
 \lambda _{1} &=&2-\frac{5}{3}\varepsilon  \nonumber \\
 \lambda _{2} &=&2-\frac{2}{3}\varepsilon  \label{105} \\
 \lambda _{3} &=&2+\frac{(n-14)}{3(n-2)}\varepsilon  \nonumber
 \end{eqnarray}
 
 \noindent and the scaling fields
 
 \begin{eqnarray}
 u_{1} &=&\delta m_{R}+\frac{2}{\left( n-1\right) }\delta (m_{A}-m_{R})-\frac{
 2}{\left( n-1\right) }\delta \left( \frac{m_{A}-m_{L}}{n}\right)  \nonumber
 \\
 u_{2} &=&-\frac{2(n-3)}{(n-1)(n-2)}\delta (m_{A}-m_{R})+\frac{4}{(n-1)}
 \delta \left( \frac{m_{A}-m_{L}}{n}\right)  \label{106} \\
 u_{3} &=&-\frac{2}{(n-1)(n-2)}\delta (m_{A}-m_{R})-\frac{2}{(n-1)}\delta
 \left( \frac{m_{A}-m_{L}}{n}\right)  \nonumber
 \end{eqnarray}
 
 \noindent where $\delta m_{i}=m_{i}-m^{*}$, $i=R,A,L$. Similar results, with 
 $n=0$, are obtained from the linearization of the mass equations, Eqs.
 (75)-(77), around the zero-field f.p. The field $u_{1}$ represents the
 average mass, i.e. $u_{1}=\left[ n\right. (n-3)/2\delta m_{R}+(n-1)\delta
 m_{A}+\left. \delta m_{L}\right] /\left[ n\right. (n-1)/\left. 2\right] $,
 while the fields $u_{2}$ and $u_{3}$ are determined by the anisotropy
 variables, $(m_{A}-m_{R})$ and $(m_{A}-m_{L})/n$. One has that
 
 \begin{eqnarray}
 u_{1} &\simeq &(t-t_{c})  \nonumber \\
 u_{2} &\simeq &2Q  \label{107} \\
 u_{3} &\simeq &-3Q^{2}.  \nonumber
 \end{eqnarray}
 
 \noindent So, we identify the critical exponent for the correlation length,
 
 \begin{equation}
 \nu =\lambda _{1}^{-1}=\frac{1}{2}+\frac{5}{12}\varepsilon  \label{108}
 \end{equation}
 
 \noindent and have the crossover exponents, for $n=0$,
 
 \begin{eqnarray}
 \phi _{2} &=&\lambda _{2}\nu =1+\allowbreak \frac{1}{2}\varepsilon
 \label{109} \\
 \phi _{3} &=&\lambda _{3}\nu =1+\allowbreak 2\varepsilon ,  \nonumber
 \end{eqnarray}
 
 \noindent which turn out to be, to order $\varepsilon $,\ the critical
 exponent for the order parameter $\beta =\phi _{2}$, and
 the critical exponent for the specific heat $\alpha =-\phi _{3}$.
 Here, $\nu $, $\beta $ and $\alpha $ are the zero-field critical exponents.
 
 The masses are given in terms of the scaling fields by
 
 \begin{eqnarray}
 \delta m_{R} &=&u_{1}+u_{2}+u_{3}  \nonumber \\
 \delta m_{A} &=&u_{1}+\frac{1}{2}\left( 4-n\right) u_{2}+(3-n)u_{3}
 \label{110} \\
 \delta m_{L} &=&u_{1}+(2-n)u_{2}+\frac{1}{2}(2-n)(3-n)u_{3}  \nonumber
 \end{eqnarray}
 
 \noindent and
 
 \begin{equation}
 \delta \left( \frac{m_{A}-m_{L}}{n}\right) =\frac{1}{2}\left[
 u_{2}+(3-n)u_{3}\right] .  \label{111}
 \end{equation}
 When $n$\ is finite, $m_{A}$\ and $m_{L}$, as well as $(m_{A}-m_{L})/n=\bar{m
 }_{AL}$,\ will grow under the renormalization group iteration, and we have
 then $m_{A}\neq m_{L}$, with different regimes occurring, according to the
 relative amplitude of the two masses. However, for $n=0$, one has that 
 $m_{A}=m_{L}$, even though $\bar{m}_{AL}$\ grows, and that single mass
 controls the behavior of the system.
 
 We now concentrate on the set of renormalization group equations with $n=0$,
 Eqs. (75)-(88), to investigate the fixed-points in a finite magnetic field 
 $h\neq 0$. One has essentially two masses, $m_{R}$ and $m_{A}=m_{L}$. On the
 AT-line, the bare masses take the values, $m_{R}=0$ and $m_{A}=m_{L}\simeq
 h^{2/3}$. Fixing the replicon modes critical, as characteristic of the spin
 glass transition, we have that the anomalous and longitudinal
 modes will scale out of the problem, 
 i.e. their mass 
  diverges under the renormalization group transformation. After a few
 iterations, we effectively obtain the reduced set of renormalization
 equations
 
 \begin{eqnarray}
 \frac{\partial m_{R}}{\partial l} &=& (2-\eta_{R})m_{R}-\left[4g_{1}^{2}-
 8g_{1}g_{2}+\frac{11}{4}g_{2}^{2}\right]I_{RR}
 \\
 \frac{\partial g_{1}}{\partial l} &=& \frac{1}{2}(\varepsilon -3 \eta_{R})g_{1}
 +\left[14g_{1}^{3}-18g_{1}^{2}g_{2}+\frac{9}{2}g_{1}g_{2}^{2}+\frac{1}{8}
 g_{2}^{3}\right]I_{RRR}
 \\
 \frac{\partial g_{2}}{\partial l} &=& \frac{1}{2}(\varepsilon -3 \eta_{R})g_{2}
 +\left[24g_{1}^{2}g_{2}-30g_{1}g_{2}^{2}+\frac{17}{2}g_{2}^{3}\right]
 I_{RRR}
 \end{eqnarray}
 with
 \begin{equation}
 \eta_{R}=\frac{1}{3}\left[4g_{1}^{2}-8g_{1}g_{2}+\frac{11}{4}g_{2}^{2}
 \right]T_{RR}
 \end{equation}
 which are equivalent to the ones studied by Bray and Roberts.$^{19}$ We 
 find the same type of fixed-points as in their work:
 
 (i) $g_{1}^{*}=0,$ $g_{2}^{*}=0$. This is the usual Gaussian f.p. which is
 unstable for $\varepsilon >0$;
 
 (ii) $g_{1}^{*2}=-\varepsilon /24,$ $g_{2}^{*}=0$. This f.p. is stable but
 unphysical, because it is complex and hence inaccessible from the domain of
 physical initial conditions via the renormalization group equations;
 
 (iii) $g_{1}^{*}=(0.415\pm i 0.090)\sqrt{\varepsilon}$, 
 $g_{2}^{*}=(0.708\mp i0.018)\sqrt{\varepsilon}$; 
 $g_{1}^{*}= i0.010\sqrt{\varepsilon}$, $g_{2}^{*}=i0.283\sqrt{\varepsilon}$.
  These 
 fixed-points are unphysical, and also turn out to be unstable.
 
 Hence no physical stable fixed-point is found to describe  the AT
 transition.
 
 We present now a set of fixed-points which we found for Eqs.(75)-(88), 
 when searching for fixed-points with critical masses $m_{R}$
 and  $m_{A}=m_{L}$, i.e., by dropping $m_{R}$ and $m_{A}$ in the 
 denominators of those equations. In addition to the zero-field
 fixed-point, discussed above, which has the symmetry
 $m_{R}^{*}=m_{A}^{*}=m_{L}^{*}$, $\bar{m}_{AL}^{*}=0$, we find
 the other following ones:
 
 (I) $g_{1}^{*}=\frac{1}{2}\sqrt{\varepsilon }$, $g_{2}^{*}=\sqrt{\varepsilon 
 }$, $g_{5}^{*}=-\frac{1}{4}\sqrt{\varepsilon }$, 
 $g_{3}^{*}=g_{6}^{*}=\bar{g}_{7}^{*}=0$, 
 $\bar{g}_{4}^{*}=-2u\sqrt{\varepsilon }$, 
 $\bar{g}_{8}^{*}=\frac{9}{4}u\sqrt{\varepsilon }$; $m_{R}^{*}=m_{A}^{*}=0$, 
 $\bar{m}_{AL}^{*}=\frac{3}{8}\varepsilon $. This in fact represents a line of
 equivalent fixed-points, $u$ being an arbitrary real number, that
 parametrizes the line. The eigenvalues, both in the coupling and mass
 spaces, remain unchanged along the line. Those fixed-points are stable in
 two directions, unstable in two directions, and marginal in one direction in
 the coupling-space, see Table II. In the mass-space the
 eigenvalues are, $\lambda _{1}=2-\frac{1}{2}\varepsilon $, $\lambda
 _{2}=\lambda _{3}=2$. These fixed-points are in fact related
 to a fixed-point that is obtained for a different symmetry 
 ($m_{R}=m_{A}=0$, $m_{L}\neq 0$), which will be discussed in a separate
 publication.$^{20}$

 (II) $g_{1}^{*}=g_{2}^{*}=g_{5}^{*}=0$, $g_{3}^{*}=-0.350\sqrt{\varepsilon }$
 , $\bar{g}_{4}^{*}=0.656\sqrt{\varepsilon }$, $g_{6}^{*}=0.519\sqrt{
 \varepsilon },$ $\bar{g}_{7}^{*}=-0.308\sqrt{\varepsilon }$, $\bar{g}
 _{8}^{*}=0.067\sqrt{\varepsilon }$; $m_{R}^{*}=-0.153\varepsilon $, 
 $m_{A}^{*}=-0.073\varepsilon $, $\bar{m}_{AL}^{*}=0.072
 \varepsilon $. This f.p. is stable in five directions and
 unstable in three directions in the coupling-space, see Table III. The
 eigenvalues in the mass-space are, $\lambda _{1}=2-0.795\varepsilon ,$ 
 $\lambda _{2}=2-0.524\varepsilon ,$ $\lambda _{3}=2-0.151\varepsilon $.
 
 (III) $g_{1}^{*}=-0.031\sqrt{\varepsilon }$, $g_{2}^{*}=0.540\sqrt{
 \varepsilon }$, $g_{3}^{*}=0.311\sqrt{\varepsilon }$, $\bar{g}_{4}^{*}=0.237\sqrt{
 \varepsilon }$, $g_{5}^{*}=0.214\sqrt{\varepsilon }$, $g_{6}^{*}=-0.368\sqrt{
 \varepsilon }$, $\bar{g}_{7}^{*}=0.134\sqrt{\varepsilon }$, $\bar{g}_{8}^{*}=0.060\sqrt{
 \varepsilon }$; $m_{R}^{*}=-0.068\varepsilon $, $m_{A}^{*}=\allowbreak
 -0.057\varepsilon $, $\bar{m}_{AL}^{*}=0.002\varepsilon $. This f.p. is
 stable in five directions and unstable in three directions in the
 coupling-space, see Table IV. The eigenvalues in the mass-space are, 
 $\lambda _{1}=2-0.519\varepsilon ,$ $\lambda _{2}=2-0.325\varepsilon ,$ 
 $\lambda _{3}=2-0.096\varepsilon $.
 
 So, although one can find different real fixed-points involving 
 $m_{R}$ and $m_{A}=m_{L}$, none is fully stable in coupling-space,
 excluding the zero-field f.p. discussed above. Each of the three
 fixed-points above attracts its own critical manifold which is spanned by
 the eigenvectors associated to the irrelevant eigenvalues ($\lambda <0$),
 i.e. the stable directions. The initial conditions in a field, correspond in
 the coupling-space to a vector where all the components are finite, see Eqs.
 (55)-(56). Comparing the structure of the critical manifolds associated to
 each of the fixed points (I), (II) and (III), as given in
 respectively Table II, III and IV, with the structure of the initial
 conditions, one may say the following. F.p. (I) is quite disconnected, f.p.
 (II) has some connection, f.p. (III) is the most connected one to the
 initial conditions. Fixed-point (II) suggests a separation of the
 coupling-space into two sectors containing, respectively, $g_{1}$, $g_{2}$, 
 $g_{5}$\ and $g_{3}$, $\bar{g}_{4}$, $g_{6}$, $\bar{g}_{7}$, $\bar{g}_{8}$,
 neither of those sectors being however completely stable. Fixed-point (III)\
 is the most likely to have some influence on the flows in the coupling-space
 in a field. Notice that the dimension of its critical manifold is equal to
 the number of distinct unbarred couplings ($g_{j}$, $j=1,2,3,5,6$). It is
 yet difficult to give a physical meaning to those three fixed points, and in
 particular connect them to a spin glass transition in a field. They may
 reflect different symmetries that are included in the general field theory
 considered.
 
 \bigskip
 
 \section{CONCLUSIONS}

 We present the general field theory, appropriate to study the transition of
 a short-range spin glass in a magnetic field, which contains three masses
 and eight couplings, explicitly written in terms of the fields for the
 replicon, anomalous and longitudinal modes. This theory allows a standard
 perturbation expansion, using propagators which involve projector operators.
 The structure of the masses and the couplings depends on the mean field
 value of the order parameter, and the number of replicas $n$. We discuss the
 symmetry of the theory in the limit $n\rightarrow 0$, and consider the
 regular case where there is a degeneracy between the anomalous and
 longitudinal masses, and a degeneracy in the couplings involving the
 anomalous and longitudinal modes. We calculate the equation of state, the
 bare masses and the bare couplings in mean-field theory. The mean-field
 calculation shows a transition in zero field, where all the modes become
 critical, and a transition in nonzero field, along the AT-line, with only
 the replicon mode critical. We then study the problem using the
 renormalization group, to order $\varepsilon =6-d$. The renormalization
 group transformation preserves the degeneracy between the anomalous and
 longitudinal masses, and the degeneracy in the couplings involving the
 anomalous and longitudinal modes, when $n$\ is fixed to zero.
 Within the general field theory we find a fixed-point associated to the
 transition in zero-field, which provides the zero-field critical exponents.
 This fixed-point which is stable in zero-field becomes unstable in the
 presence of a small magnetic field, and we calculate crossover exponents,
 which we relate to zero-field critical exponents. It is, however, not clear
 where the crossover leads to. For a finite magnetic field, we find no
 physical stable fixed-point, that would describe the AT transition, in
 agreement with the results of Bray and Roberts.$^{19}$ The absence of stable
 fixed-points, accessible from the domain of physical initial Hamiltonians,
 can have different interpretations in the renormalization group, as
 discussed in Ref. 19: the fluctuations may drive the transition first order,
 the transition may be first order even within mean-field theory, or the
 fluctuations may destroy the transition. To conclude, the present study
 shows that a theory with degeneracy between the anomalous and longitudinal
 masses, as imposed by keeping $n$ equal zero,\ leaves no place for a second
 order spin glass transition outside zero-field.

 \section*{ACKNOWLEDGMENTS}

 We are grateful to E. Br\'{e}zin for helpful discussions. This work has
 been supported by the Hungarian Science Fund (OTKA), grant No. T032424.
 
 \bigskip
 
 \appendix
 \section*{APPENDIX}
 
 We present here the derivation of the expressions for the longitudinal,
 anomalous and replicon projectors given in Eqs. (18), (25) and (30). We
 obtain those expressions following a procedure similar to the one used by
 Bray and Roberts to get the components of the replicon projector in their
 work.$^{19}$
 
 The longitudinal projector is defined by, Eq. (17),
 
 \[
 \hspace{1.2in}
 P^{L}=e^{L}e^{L}  
 \hspace{2.7in}(A1)
 \]
 where $e^{L}$\ represents a component of the replica independent unit vector,
 $\mathbf{e}^{L}\equiv \sqrt{2/n(n-1)}\left[ 1,\ldots ,1\right] $,
 
 \[
 \hspace{1.2in}
 e_{\alpha \beta }^{L}=e^{L}  
 \hspace{2.85in}(A2)
 \]
 
 \noindent The longitudinal projector has then only one component, which is
 obviously given by
 
 \[
 \hspace{1.2in}
 P^{L}=\frac{2}{n(n-1)}. 
 \hspace{2.43in}(A3)
 \]
 
 The anomalous projector is defined by, Eq. (26),
 
 \[
 \hspace{1.2in}
 P_{\alpha ,\beta }^{A}={\sum_{\mu =1}^{n-1}}e_{\alpha }^{A,\mu
 }e_{\beta }^{A,\mu }  
 \hspace{2.24in}(A4)
 \]
 
 \noindent where the vectors $\mathbf{e}^{A,\mu }$ have the property
 
 \[
 \hspace{1.2in}
 {\sum_{\alpha } }e_{\alpha }^{A,\mu }=0  
 \hspace{2.76in}(A5)
 \]
 
 \noindent and normalization ${\sum_{\alpha } }e_{\alpha }^{A,\mu
 }e_{\alpha }^{A,\mu }=\frac{4}{(n-2)}$ (which follows from the fact that 
 $\mathbf{e}^{A,\mu }$ is a unit vector in the space of replica pairs and 
 $e_{\alpha \beta }^{A,\mu }=\frac{1}{2}(e_{\alpha }^{A,\mu }+e_{\beta
 }^{A,\mu })$ ). The anomalous projector has two distinct components: 
 $P_{\alpha ,\alpha }^{A}$ and $P_{\alpha ,\beta }^{A}$ with $\alpha \neq
 \beta $. Setting $\alpha =\beta $ in Eq. (A4) and summing both sides over 
 $\alpha $, gives
 
 \begin{eqnarray*}
 \hspace{1.2in}
 nP_{\alpha ,\alpha }^{A} &=&{\sum_{\mu =1}^{n-1}}
 {\sum_{\alpha } }e_{\alpha }^{A,\mu }e_{\alpha }^{A,\mu }  \nonumber \\
 &=&\frac{4(n-1)}{(n-2)};  
 \hspace{2.1in}(A6)
 \end{eqnarray*}
 
 \noindent setting $\alpha \neq \beta $ in Eq. (A4) and summing both sides
 over $\alpha \neq \beta $, gives
 
 \begin{eqnarray*}
 \hspace{1.2in}
 n(n-1)P_{\alpha ,\beta }^{A} &=&{\sum_{\mu =1}^{n-1}}
 {\sum_{\alpha \neq \beta } }e_{\alpha }^{A,\mu }e_{\beta }^{A,\mu }  \nonumber
 \\
 &=&-{\sum_{\mu =1}^{n-1}}{\sum_{\alpha } }e_{\alpha
 }^{A,\mu }e_{\alpha }^{A,\mu }  \nonumber \\
 &=&-\frac{4(n-1)}{(n-2)},  
 \hspace{1.5in}(A7)
 \end{eqnarray*}
 
 \noindent where we used Eq. (A5). Eqs. (A6) and (A7) lead to the general
 form for the anomalous projector,
 
 \[
 \hspace{1.2in}
 P_{\alpha ,\beta }^{A}=\frac{4}{(n-2)}\left( \delta _{\alpha \beta }-\frac{1
 }{n}\right)  
 \hspace{1.75in}(A8)
 \]
 
 The replicon projector is defined by, Eq. (31),
 
 \[
 \hspace{1.2in}
 P_{\alpha \beta ,\gamma \delta }^{R}={\sum_{\nu =1}^{\frac{1}{2}
 n(n-3)}}e_{\alpha \beta }^{R,\nu }e_{\gamma \delta }^{R,\nu }.  
 \hspace{1.85in}(A9)
 \]
 
 \noindent where $\mathbf{e}^{R,\nu }$ are unit vectors, which have the
 property
 
 \[
 \hspace{1.2in}
 {\sum_{\alpha (\neq \beta )} }e_{\alpha \beta }^{R}=0.  
 \hspace{2.58in}(A10)
 \]
 
 \noindent The replicon projector $P_{\alpha \beta ,\gamma \delta }^{R}$ has
 three different components: $P_{\alpha \beta ,\alpha \beta }^{R}$, 
 $P_{\alpha \beta ,\alpha \delta }^{R}$ with $\beta \neq \delta $, and 
 $P_{\alpha \beta ,\gamma \delta }^{R}$ with $\alpha \neq \gamma ,\beta \neq
 \delta $. Setting $\alpha =\gamma $, $\beta =\delta $ in Eq. (A9) and
 summing both sides over distinct $\alpha ,\beta ,$ gives
 
 \begin{eqnarray*}
 \hspace{1.2in}
 n(n-1)P_{\alpha \beta ,\alpha \beta }^{R} &=&-{\sum_{\nu =1}^{\frac{
 1}{2}n(n-3)}}{\sum_{\alpha \neq \beta } }e_{\alpha \beta }^{R,\nu
 }e_{\alpha \beta }^{R,\nu }  \nonumber \\
 &=&n(n-3);  
 \hspace{1.41in}(A11)
 \end{eqnarray*}
 setting $\alpha =\gamma $, $\beta \neq \delta $ in Eq. (A9) and summing
 both sides over distinct $\alpha $, $\beta $, $\delta $ , gives
 
 \begin{eqnarray*}
 \hspace{1.2in}
 n(n-1)(n-2)P_{\alpha \beta ,\alpha \delta }^{R} &=&{\sum_{\nu =1}
 ^{\frac{1}{2}n(n-3)}}{\sum_{\alpha ,\beta ,\delta } }^{\prime}
 e_{\alpha \beta }^{R,\nu }e_{\alpha \delta }^{R,\nu }  \nonumber \\
 &=&-{\sum_{\nu =1}^{\frac{1}{2}n(n-3)}}
 {\sum_{\alpha \neq\beta } }e_{\alpha \beta }^{R,\nu }
 e_{\alpha \beta }^{R,\nu }  \nonumber
 \\
 &=&-n(n-3)  
 \hspace{0.89in}(A12)
 \end{eqnarray*}
 
 \noindent where we used Eq. (A10); setting $\alpha \neq \gamma $, $\beta
 \neq \delta $ in Eq. (A9) and summing both sides over distinct $\alpha $, 
 $\beta $, $\gamma $, $\delta $ , gives, using again Eq. (A10),
 
 \begin{eqnarray*}
 \hspace{0.7in}
 n(n-1)(n-2)(n-3)P_{\alpha \beta ,\gamma \delta }^{R} &=&
 {\sum_{\nu =1}^{\frac{1}{2}n(n-3)}}
 {\sum_{\alpha ,\beta ,\gamma ,\delta } }
 ^{\prime }e_{\alpha \beta }^{R,\nu }e_{\gamma \delta }^{R,\nu }  \nonumber \\
 &=&-{\sum_{\nu =1}^{\frac{1}{2}n(n-3)}}
 {\sum_{\alpha ,\beta,\delta } }^{\prime }
 e_{\alpha \beta }^{R,\nu }(e_{\alpha \delta
 }^{R,\nu }+e_{\beta \delta }^{R,\nu })  \nonumber \\
 &=&2n(n-3)  
 \hspace{1.in}(A13)
 \end{eqnarray*}
 
 \noindent Eqs. (A11), (A12) and (A13) lead to the general form for the
 replicon projector,
 
 \begin{eqnarray*}
 \hspace{0.2in}
 P_{\alpha \beta ,\gamma \delta }^{R} &=&\left( \delta _{\alpha \gamma
 }\delta _{\beta \delta }+\delta _{\alpha \delta }\delta _{\beta \gamma
 }\right) -\left( \delta _{\alpha \gamma }+\delta _{\alpha \delta }+\delta
 _{\beta \gamma }+\delta _{\beta \delta }\right) \frac{1}{n-2} \\
 &&+\frac{2}{(n-1)(n-2)}
 \hspace{2.55in}(A14)
 \end{eqnarray*}
 
 \noindent with $\alpha \neq \beta $, $\gamma \neq \delta $. The results in
 Eqs. (A.11), (A.12) and (A.13), are similar to the ones obtained by Bray and
 Roberts.$^{19}$
 
 \newpage
 
 \section*{References}

 \noindent $^{1}$K. Binder and A.P. Young, Rev. Mod. Phys. \textbf{58}, 801
 (1986).
 
 \noindent $^{2}$M. M\'{e}zard, G. Parisi, and M.A. Virasoro, \textit{Spin
 Glass Theory and Beyond} (World Scientific, Singapore, 1987).
 
 \noindent $^{3}$K.H. Fischer and J.H. Hertz, \textit{Spin Glasses}
 (Cambridge University Press, Cambridge, 1991).
 
 \noindent $^{4}$\textit{Spin Glasses and Random Fields}, edited by A.P.
 Young (World Scientific, Singapore, 1998).
 
 \noindent $^{5}$\textit{Spin Glasses and Biology}, edited by D.\ Stein
 (World Scientific, Singapore 1992).
 
 \noindent $^{6}$S.F. Edwards and P.W. Anderson, J. Phys. F \textbf{5}, 965
 (1975).
 
 \noindent $^{7}$G. Parisi, Phys. Rev. Lett. \textbf{43}, 1754 (1979); J.
 Phys. A \textbf{13}, L115,\textbf{13}, 1101, \textbf{13}, 1887 (1980); 
 Phys. Rev. Lett. \textbf{50}, 1946 (1983).
 
 \noindent $^{8}$D. Sherrington and S. Kirkpatrick, Phys. Rev. Lett. \textbf{
 35}, 1792 (1975).
 
 \noindent $^{9}$D.S. Fisher and D.A. Huse, Phys. Rev. Lett. \textbf{56}, 1601
 (1986); Phys. Rev. B \textbf{38}, 386 (1988).
 
 \noindent $^{10}$W.L. McMillan, J. Phys. \textbf{C} 17, 3179 (1984).
 
 \noindent $^{11}$A. J. Bray and M.A. Moore, in \textit{Proceedings of the
 Heidelberg Colloquium on Glassy Dynamics}, edited by J.L. van Hemmen and I.
 Morgenstern (Springer Verlag, Heidelberg, 1986),p. 121; 
 Phys. Rev. Lett. \textbf{58}, 57 (1987).
 
 \noindent $^{12}$J.R.L. de Almeida and D.J.\ Thouless, J. Phys. A \textbf{11}
 , 983 (1977).
 
 \noindent $^{13}$A.J. Bray and M.A. Moore, J. Phys. C \textbf{12}, 79 (1979).
 
 \noindent $^{14}$P. Nordblad and P. Svedlindh, in \textit{Spin Glasses and
 Random Fields}, edited by A.P. Young (World Scientific, Singapore, 1998),p. 1.
 
 \noindent $^{15}$P. Monod and H. Bouchiat, J. Physique Lett. \textbf{43},
 L45 (1982); V.S. Zotev and R. Orbach,(submitted), cond-mat/0201226.
 
 \noindent $^{16}$E. Marinari, G. Parisi, and J.J. Ruiz-Lorenzo, in \textit{
 Spin Glasses and Random Fields}, edited by A.P. Young (World Scientific,
 Singapore, 1998),p. 59; J. Houdayer and O.C. Martin, Phys. Rev. Lett.
 \textbf{82}, 49, 34 (1999); F. Krzakala, J. Houdayer, E. Marinari, O.
 C. Martin, and G. Parisi, Phys. Rev. Lett. \textbf{87}, 197204 (2001).
 
 \noindent $^{17}$J.E. Green, M.A. Moore, and A.J. Bray, J. Phys. C \textbf{16
 }, L815 (1983).
 
 \noindent $^{18}$A.B. Harris, T.C. Lubensky, and J-H. Chen, Phys. Rev. Lett. 
 \textbf{36}, 415 (1976).
 
 \noindent $^{19}$A.J. Bray and S.A. Roberts, J. Phys. C \textbf{13}, 5405
 (1980).
 
 \noindent $^{20}$T. Temesvari, C. De Dominicis, and I.R. Pimentel, 
 (submitted).
 
 \noindent $^{21}$T. Temesvari, C. De Dominicis, and I.R. Pimentel, Eur.
 Phys. J. B \textbf{25}, 361 (2002).
 
 \noindent $^{22}$C. De Dominicis, D.M. Carlucci, and T. Temesv\'{a}ri, J.
 Phys. I France \textbf{7}, 105 (1997).
 
 \noindent $^{23}$E. Pytte and J. Rudnick, Phys.Rev. B \textbf{19}, 3603
 (1979).
 
 \noindent $^{24}$K.G. Wilson and J.B. Kogut, Phys. Rep. 12C, 75 (1974).

 \newpage
 
 \noindent

 TABLE\ I. Eigenvalues $\lambda $ and eigenvectors $E$, in coupling-space for
 the zero-field f.p..
 
 \bigskip
 \bigskip
 
 \begin{tabular}{cccc}
 \hline\hline
 $\lambda $ & $-1$ & $-3.82$ & $0.74\pm i2.54\medskip $ \\ 
 $E$ & $\left[ 
 \begin{array}{c}
 1 \\ 
 0 \\ 
 -1 \\ 
 0 \\ 
 -1 \\ 
 2 \\ 
 -3/2 \\ 
 1/4
 \end{array}
 \right] $ & $\left[ 
 \begin{array}{c}
 0.51 \\ 
 0.22 \\ 
 -0.26 \\ 
 -0.84 \\ 
 -0.48 \\ 
 1 \\ 
 -0.75 \\ 
 0.19
 \end{array}
 \right] $ & $\left[ 
 \begin{array}{c}
 \allowbreak 0.09\pm i0.06 \\ 
 -0.50\pm i0.36 \\ 
 -0.50\pm i0.12 \\ 
 \allowbreak 0.34\pm i0.05 \\ 
 -0.41\pm i0.01 \\ 
 1 \\ 
 -0.79\pm i0.10 \\ 
 \allowbreak 0.22\pm i0.07
 \end{array}
 \right] \medskip $ \\ 
 $\lambda $ & $1$ & $7$ & $(10\pm i\medskip \sqrt{23})/3$ \\ 
 $E$ & $\medskip \left[ 
 \begin{array}{c}
 3/7 \\ 
 -16/7 \\ 
 -2 \\ 
 1 \\ 
 -11/7 \\ 
 26/7 \\ 
 -41/14 \\ 
 45/56
 \end{array}
 \right] $ & $\allowbreak \left[ 
 \begin{array}{c}
 1/5 \\ 
 -4/5 \\ 
 -1 \\ 
 1 \\ 
 -1 \\ 
 3 \\ 
 -3 \\ 
 3/2
 \end{array}
 \right] $ & $\left[ 
 \begin{array}{c}
 (13\pm i\sqrt{23})/64 \\ 
 -(19\mp i\sqrt{23})/16 \\ 
 -(59\mp i\sqrt{23})/48 \\ 
 1 \\ 
 -(205\pm i\sqrt{23})/192 \\ 
 3 \\ 
 -(85\pm i\sqrt{23})/32 \\ 
 (135\pm i\sqrt{207})/128
 \end{array}
 \right] $ \\ \hline\hline
 \end{tabular}
 \bigskip

 \newpage
 
 TABLE II. Eigenvalues $\lambda $ and eigenvectors $E$, in coupling-space for
 f.p. (I), \{degeneracy\}.
 
 \bigskip
 
 \bigskip $
 \begin{tabular}[t]{ccccc}
 \hline\hline
 $\lambda $ & $-1$ & $(2\pm \sqrt{91})/6$ & $0$ $\left\{ 2\right\} $ & $1/2$ $
 \left\{ 3\right\} $ \\ 
 $E$ & $\left[ 
 \begin{array}{c}
 1 \\ 
 2 \\ 
 0 \\ 
 0 \\ 
 -1/2 \\ 
 0 \\ 
 0 \\ 
 0
 \end{array}
 \right] $ & $\left[ 
 \begin{array}{c}
 1 \\ 
 (14\pm 2\sqrt{91})/3 \\ 
 0 \\ 
 0 \\ 
 (11\pm \sqrt{91})/6 \\ 
 0 \\ 
 0 \\ 
 0
 \end{array}
 \right] \medskip $ & $\left[ 
 \begin{array}{c}
 0 \\ 
 0 \\ 
 0 \\ 
 1 \\ 
 0 \\ 
 0 \\ 
 0 \\ 
 -9/8
 \end{array}
 \right] $ & $\left[ 
 \begin{array}{r}
 0 \\ 
 0 \\ 
 0 \\ 
 0 \\ 
 0 \\ 
 0 \\ 
 0 \\ 
 1
 \end{array}
 \right] $ \\ \hline\hline
 \end{tabular}
 $

 \newpage 
 
 TABLE III. Eigenvalues $\lambda $ and eigenvectors $E$, in coupling-space
 for f.p. (II).
 \bigskip
 \bigskip
 
 \begin{tabular}{ccccc}
 \hline\hline
 $\lambda $ & $0.606$ & $-1.692$ & $-1.001\medskip $ & $-0.319\pm
 i1.818$ \\ 
 $E$ & $\left[ 
 \begin{array}{c}
 0 \\ 
 0 \\ 
 0.83 \\ 
 0.41 \\ 
 0 \\ 
 0.55 \\ 
 -1.00 \\ 
 -0.64
 \end{array}
 \right] $ & $\left[ 
 \begin{array}{c}
 0 \\ 
 0 \\ 
 0.90 \\ 
 0.24 \\ 
 0 \\ 
 0.84 \\ 
 1.00 \\ 
 0.48
 \end{array}
 \right] $ & $\left[ 
 \begin{array}{c}
 0 \\ 
 0 \\ 
 1.00 \\ 
 0.32 \\ 
 0 \\ 
 0.68 \\ 
 0.89 \\ 
 0.42
 \end{array}
 \right] $ & $\left[ 
 \begin{array}{c}
 0 \\ 
 0 \\ 
 -0.13\pm i1.08 \\ 
 -0.21\pm i0.39 \\ 
 0 \\ 
 -0.32\pm i0.80 \\ 
 -1.00\pm i1.02 \\ 
 -0.73\pm i0.25
 \end{array}
 \right] \medskip $ \\ 
 $\lambda $ & $-1.793$ & $0.895$ & $0.002\medskip $ &  \\ 
 $E$ & $\medskip \left[ 
 \begin{array}{c}
 0.63 \\ 
 -0.39 \\ 
 0 \\ 
 0 \\ 
 1.00 \\ 
 0 \\ 
 0 \\ 
 0
 \end{array}
 \right] $ & $\left[ 
 \begin{array}{c}
 1.00 \\ 
 -0.02 \\ 
 0 \\ 
 0 \\ 
 0.34 \\ 
 0 \\ 
 0 \\ 
 0
 \end{array}
 \right] $ & $\left[ 
 \begin{array}{c}
 -0.79 \\ 
 0.57 \\ 
 0 \\ 
 0 \\ 
 1.00 \\ 
 0 \\ 
 0 \\ 
 0
 \end{array}
 \right] $ &  \\ \hline\hline
 \end{tabular}
 
 \newpage
 
 TABLE IV. Eigenvalues $\lambda $ and eigenvectors $E$, in coupling-space for
 f.p. (III).
 \bigskip
 \bigskip 
 
 \begin{tabular}{ccccc}
 \hline\hline
 $\lambda $ & $-1.45$ & $-1.00$ & $-0.65\medskip $ & $-0.37\pm
 i0.15$ \\ 
 $E$ & $\left[ 
 \begin{array}{r}
 0.03 \\ 
 -1.00 \\ 
 -0.63 \\ 
 -0.41 \\ 
 -0.48 \\ 
 0.93 \\ 
 -0.35 \\ 
 -0.14
 \end{array}
 \right] $ & $\left[ 
 \begin{array}{r}
 -0.06 \\ 
 1.00 \\ 
 0.57 \\ 
 0.44 \\ 
 0.40 \\ 
 -0.68 \\ 
 0.25 \\ 
 0.11
 \end{array}
 \right] $ & $\left[ 
 \begin{array}{r}
 0.26 \\ 
 0.08 \\ 
 0.08 \\ 
 -1.00 \\ 
 0.14 \\ 
 -0.58 \\ 
 0.70 \\ 
 -0.18
 \end{array}
 \right] $ & $\left[ 
 \begin{array}{r}
 0.54\mp i0.49 \\ 
 1.00\pm i0.51 \\ 
 -0.04\pm i0.21 \\ 
 0.65\pm i1.93 \\ 
 -0.17\pm i0.12 \\ 
 0.48\pm i0.22 \\ 
 -0.65\mp i0.56 \\ 
 0.18\pm i0.25
 \end{array}
 \right] \medskip $ \\ 
 $\lambda $ & $0.31$ & $0.16$ & $0.04\medskip $ &  \\ 
 $E$ & $\left[ 
 \begin{array}{r}
 -0.44 \\ 
 -0.70 \\ 
 0.15 \\ 
 0.33 \\ 
 0.89 \\ 
 -0.67 \\ 
 1.00 \\ 
 -0.01
 \end{array}
 \right] $ & $\left[ 
 \begin{array}{r}
 -0.04 \\ 
 0.22 \\ 
 0.16 \\ 
 -0.06 \\ 
 0.15 \\ 
 -0.70 \\ 
 0.99 \\ 
 -1.00
 \end{array}
 \right] $ & $\left[ 
 \begin{array}{r}
 0.03 \\ 
 -1.00 \\ 
 -0.60 \\ 
 0.43 \\ 
 -0.37 \\ 
 0.60 \\ 
 -0.35 \\ 
 -0.14
 \end{array}
 \right] \medskip $ &  \\ \hline\hline
 \end{tabular}
 
 \end{document}